\documentclass[a4paper,11pt]{article}
\usepackage[english]{babel}
\usepackage{jheppub}
\usepackage{subfig}
\usepackage[T1]{fontenc} 
\usepackage{graphicx}
\usepackage{epsfig}
\usepackage{axodraw}
\usepackage{amsmath}
\usepackage{amssymb}
\usepackage{float}
\usepackage{placeins}
\usepackage{xcolor}
\usepackage{shuffle}
\usepackage{mathtools}
\usepackage{mathrsfs}
\usepackage{float}

\allowdisplaybreaks[4]

\title{Three-loop contributions to the $\rho$ parameter and iterated integrals of modular forms}

\author[a]{Samuel Abreu,}
\author[a]{Matteo Becchetti,}
\author[b]{Claude Duhr,}
\author[c]{and Robin Marzucca}

\affiliation[a]{Center for Cosmology, Particle Physics and Phenomenology (CP3), Universit\'e Catholique de Louvain, 1348 Louvain-La-Neuve, Belgium}

\affiliation[b]{Theoretical Physics Department, CERN, Geneva, Switzerland}
\affiliation[c]{IPPP, Department of Physics, Durham University, Durham DH1 3LE, United Kingdom}

\emailAdd{samuel.abreu@uclouvain.be, matteo.becchetti@uclouvain.be, claude.duhr@cern.ch, robin.marzucca@durham.ac.uk}

\abstract{We compute fully analytic results for the three-loop diagrams involving two different massive quark flavours 
contributing to the $\rho$ parameter in the Standard Model. We find that the results
involve exactly the same class of functions that appears in the
well-known sunrise and banana graphs, namely elliptic polylogarithms and iterated integrals of
modular forms. Using recent developments in the understanding of these functions,
we analytically continue all the iterated integrals of modular forms to all regions of the parameter space,
and in each region we obtain manifestly real and fast-converging series expansions for these functions.
}

\newcommand{\be}{\begin{equation}}
\newcommand{\ee}{\end{equation}}
\newcommand{\nn}{\nonumber}
\newcommand{\beq}{\begin{equation}}
\newcommand{\eeq}{\end{equation}}
\newcommand{\bea}{\begin{eqnarray}}
\newcommand{\eea}{\end{eqnarray}}
\newcommand{\bfig}{\begin{figure}}
\newcommand{\efig}{\end{figure}}
\newcommand{\bc}{\begin{center}}
\newcommand{\ec}{\end{center}}

\newcommand{\eps}{\epsilon}

\newcommand{\cEf}[4]{{\mathcal{E}_3}\!\left(\begin{smallmatrix}#1\\#2\end{smallmatrix};#3,#4\right)}
\newcommand{\IEs}[2]{{I}\!\left(\begin{matrix}#1\end{matrix};#2\right)}

\DeclareMathOperator{\SL}{SL}
\DeclareMathOperator{\K}{K}
\DeclareMathOperator{\E}{E}

\begin{document}

\begin{flushright}
\preprint{CERN-TH-2019-210, CP3-19-53, IPPP/19/90}
\end{flushright}

\maketitle
\flushbottom


\section{Introduction}

Multiloop Feynman integrals are a cornerstone of perturbative Quantum Field Theory,
as they are the main mathematical building blocks for the computation of higher-order
corrections to physical observables. For this reason, a lot of effort has been put in understanding
their mathematical structure and the class of special functions to which they evaluate.
For instance, unitarity dictates that Feynman integrals must be multivalued functions
with logarithmic branch cuts, and this must be reflected in the corresponding class of functions.

A lot of progress was made over the last decade in understanding the simplest class of
special functions that appear in multiloop computations. It was realised that many Feynman integrals can be evaluated
in terms of multiple polylogarithms~\cite{goncharov2011multiple,Remiddi:1999ew,Gehrmann:2000zt}.
These functions are by now well understood, including their analytic continuation to
arbitrary kinematic regions and their efficient numerical
evaluation for arbitrary
complex arguments~\cite{Vollinga:2004sn,Buehler:2011ev,Frellesvig:2016ske,Ablinger:2018sat}.

While multiple polylogarithms are sufficient to express all one-loop integrals
in four space-time dimensions, it has been known for a long time that new classes of functions
can appear starting from two-loop order. This was noted for the first time in
the calculation of the two-loop corrections
to the electron propagator in QED with massive electrons~\cite{Sabry}.
The simplest two-loop integral which cannot be expressed in terms of multiple polylogarithms
is the so-called two-loop sunrise graph with three massive propagators, which has been extensively studied over the last
decades~\cite{Broadhurst:1987ei,Bauberger:1994by,Bauberger:1994hx,Caffo:1998du,Laporta:2004rb,Kniehl:2005bc,Caffo:2008aw,
MullerStach:2011ru,Bloch:2013tra,Remiddi:2013joa,Adams:2013kgc,Adams:2013nia,Adams:2014vja,Adams:2015gva,Adams:2015ydq,Remiddi:2016gno,Bloch:2016izu,Broedel:2017siw,Bogner:2019lfa}.
Recent interest in the sunrise graph was fuelled to a large extend by the fact that the same type of functions appears
in many higher-order calculations for the Large Hadron Collider (LHC) at CERN where
the mass of the top quark cannot be neglected, see for example refs~\cite{Czakon:2013goa,vonManteuffel:2017hms,Adams:2018kez,Adams:2018bsn,Bonciani:2016qxi,Bonciani:2019jyb,Francesco:2019yqt,Frellesvig:2019byn,Becchetti:2017abb}.
By now, several representations of the sunrise graph in terms of different classes
of special functions are known. They all have in common that they involve
functions related to elliptic curves, most prominently elliptic multiple polylogarithms and iterated integrals
of modular forms.

Since elliptic functions seem to be a feature of many two-loop diagrams
with massive propagators, it is natural to expect that
these functions also prominently show up when performing calculations in the electroweak sector of the Standard Model (SM),
where the gauge bosons and the fermions acquire a mass through the Higgs mechanism.
In this paper we consider one of the precision observables of the electroweak SM,
namely the corrections to the $\rho$ parameter, defined as the difference between the vacuum polarisations
of the $W$ and $Z$ bosons. The $\rho$ parameter is known through three-loop order
in the SM in the limit where all quarks but the top quark are massless~\cite{Veltman:1977kh,Djouadi:1987di,Chetyrkin:1995ix,Avdeev:1994db,Chetyrkin:2006bj,Boughezal:2006xk,Schroder:2005db}.
In ref.~\cite{Grigo:2012ji}, corrections from three-loop diagrams were considered
in the scenario where also the bottom quark has a non-vanishing mass. The results of ref.~\cite{Grigo:2012ji} where presented as an expansion
in the ratio of the two quark masses. However, no closed analytic formula was presented,
because the corresponding loop integrals were observed to involve functions of elliptic type,
and at the time the theory of these functions was still largely underdeveloped.

While the series expansion of ref.~\cite{Grigo:2012ji} is sufficient
to obtain reliable phenomenological predictions, it is interesting
to revisit the computation of ref.~\cite{Grigo:2012ji} in the light of
the recent developments in the understanding of elliptic Feynman integrals.
In this way one can obtain fully analytic results at high-loop order for one of the most
fundamental precision observables of the SM. First steps in this direction were taken in
refs.~\cite{Ablinger:2017bjx,Blumlein:2018aeq}, where the elliptic Feynman integrals of
ref.~\cite{Grigo:2012ji} were computed in terms of a class of new transcendental
functions, called iterative non-iterative integrals. The relationship between these functions
and the functions that appear in the different known representations of the sunrise graph
is not fully clear. It is an interesting question if completely new classes of elliptic-type
functions show up in the computation of the $\rho$ parameter which are not covered
by the existing literature on the sunrise graph and on elliptic polylogarithms and iterated integrals of modular forms.

The purpose of this paper is to show that all the Feynman integrals that contribute
to the three-loop corrections to the $\rho$ parameter with two massive
quark flavours can be expressed in terms of exactly the same class of functions that appear
in the sunrise graph. We show how all the elliptic Feynman
integrals of ref.~\cite{Grigo:2012ji} can be performed in terms of iterated integrals of modular forms
for the same congruence subgroup as for the sunrise integral.
These iterated integrals can be analytically
continued to the whole parameter space in such a way that they admit fast-converging series representations
for all values of the quark masses.

Our paper is organised as follows: In Section~\ref{sec:notAndConv} we review the results
of ref.~\cite{Grigo:2012ji} and we identify the elliptic Feynman integrals that need to be computed.
In Section~\ref{sec:3} we introduce the mathematical background on elliptic curves and elliptic
polylogarithms needed throughout the paper. In Section~\ref{sec:f28_eMPL} we present our first main result,
and we show how to evaluate the Feynman parameter representation for the simplest elliptic Feynman integral
in terms of elliptic polylogarithms. In Section~\ref{Sec5} we review the connection
between elliptic polylogarithms and iterated integrals of modular forms, and in Section~\ref{Sec6} we use this relationship
to obtain analytic results for all elliptic Feynman integrals that contribute to the three-loop $\rho$ parameter
in the region where the ratio of the quark masses is small. In Section~\ref{Sec7} we discuss the analytic continuation of these integrals
to the whole parameter space. In Section~\ref{Sec8} we present our final analytic result for the $\rho$ parameter
at three loops, and in Section~\ref{sec:conclusion} we draw our conclusions. We include several appendices where we
collect formulas omitted throughout the main text.


\section{Notations and computational setting}
\label{sec:notAndConv}

In this section we review the background to
the three-loop QCD corrections to the electroweak $\rho$ parameter with two massive quark flavours. 
We closely follow the presentation of ref.~\cite{Grigo:2012ji}. The $\rho$ parameter can be written as:
\begin{equation}
\rho = 1 + \delta \rho,
\end{equation}
where the higher-order corrections are given by
\begin{equation}
\delta \rho = \frac{\Sigma_Z(0)}{M^2_Z} - \frac{\Sigma_W(0)}{M^2_W}.
\end{equation}
$\Sigma_Z(0)$ and $\Sigma_W(0)$ are, respectively, the transverse parts of the $Z$ and $W$ boson propagators at zero momentum. They are defined as
\begin{equation}
\Sigma_{Z/W}(0) = \frac{g_{\mu\nu}}{d}\Pi^{\mu\nu}_{Z/W},
\end{equation}
where $\Pi^{\mu\nu}_{Z/W}$ are the correlator functions for the $Z$ and $W$ bosons, and $d$ is the space-time dimension.

We will consider the three-loop QCD corrections to the $\rho$ parameter in $n_f$-flavour QCD with $n_f - 2$ massless quarks and two massive ones, whose masses we  denote by $m_1$ and $m_2$. In the SM, this corresponds to $n_f=6$, and $m_1$ and $m_2$ denote the masses of the top and bottom quarks respectively. It is possible to write the higher-order corrections to the $\rho$ parameter as an expansion in the strong coupling constant $\alpha_s$:
\begin{equation}
\delta \rho = \frac{3 G_F m^2_{\textrm{t}}}{16 \pi^2 \sqrt{2}}\left(\delta^{(0)} + \frac{\alpha_s(\mu)}{\pi} \delta^{(1)} + \left(\frac{\alpha_s(\mu)}{\pi}\right)^2\delta^{(2)} + \mathcal{O}(\alpha_s(\mu)^3)\right),
\end{equation}
where $G_F$ is the Fermi constant and $m_{\textrm{t}}$ is the top-quark $\overline{\textrm{MS}}$-mass.
We will choose the renormalisation scale as $\mu^2 = m_{\textrm{t}}^2$. 
The term $\delta^{(j)}$ involves graphs with $j+1$ loops, and this paper is devoted to the calculation
of certain three-loop masters appearing in $\delta^{(2)}$.

\begin{figure}
\renewcommand*\thesubfigure{\arabic{subfigure}} 
\centering
\subfloat[]{\includegraphics[width =2 cm]{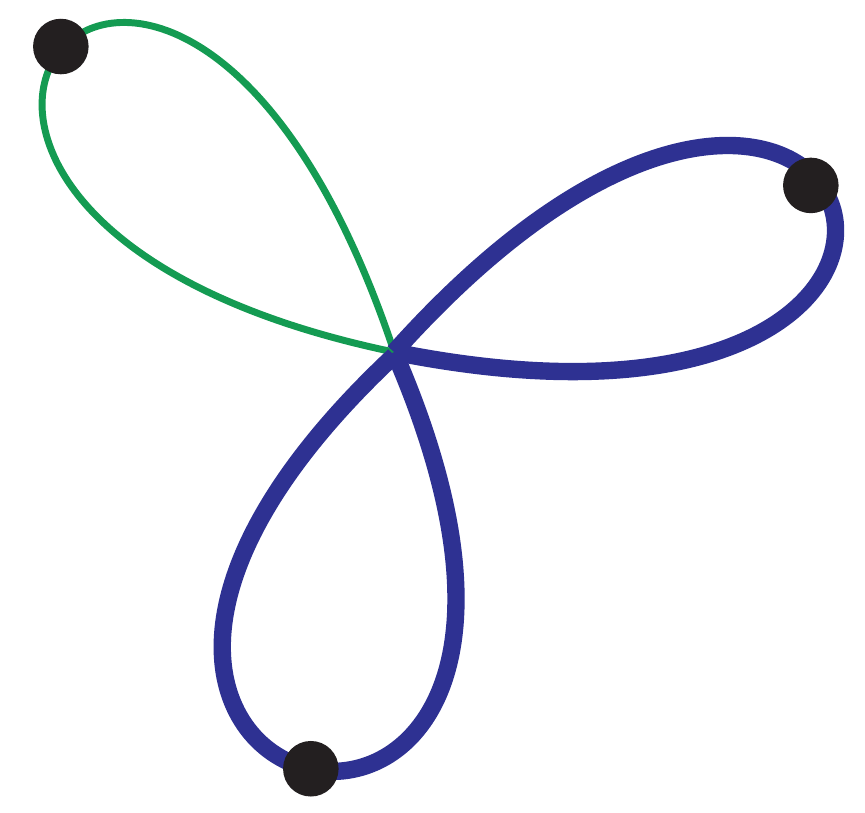}}\qquad
\subfloat[]{\includegraphics[width =2 cm]{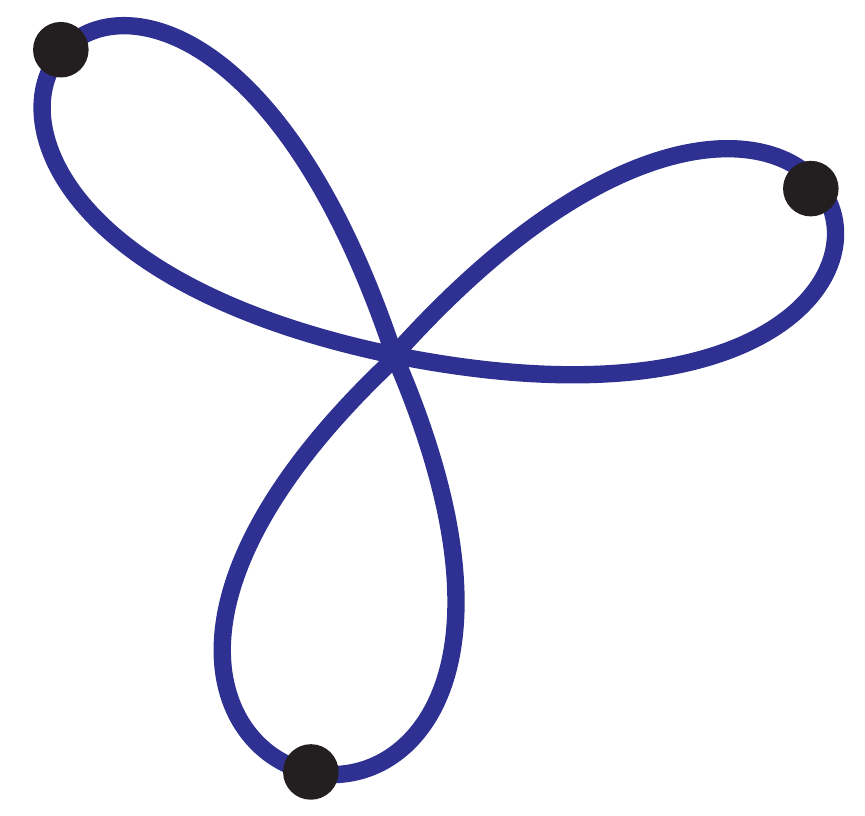}}\qquad
\subfloat[]{\includegraphics[width =2 cm]{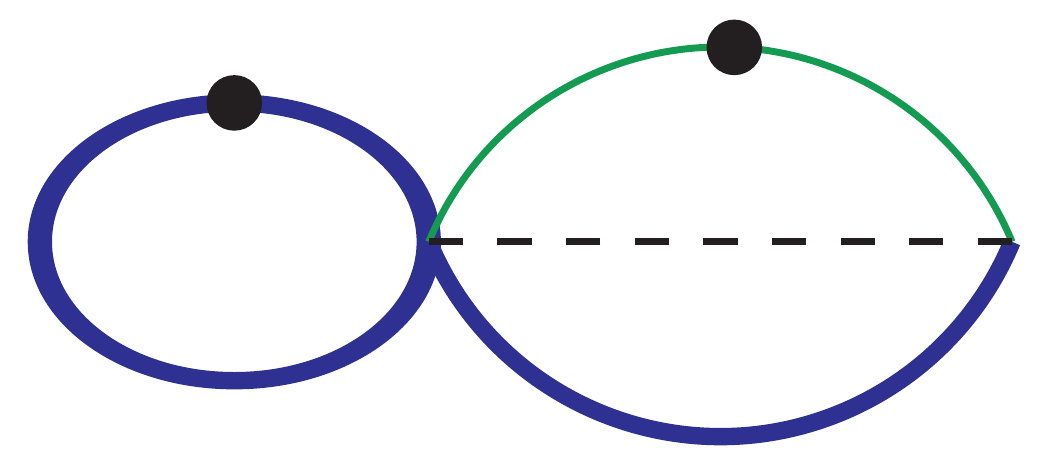}}\qquad
\subfloat[]{\includegraphics[width =2 cm]{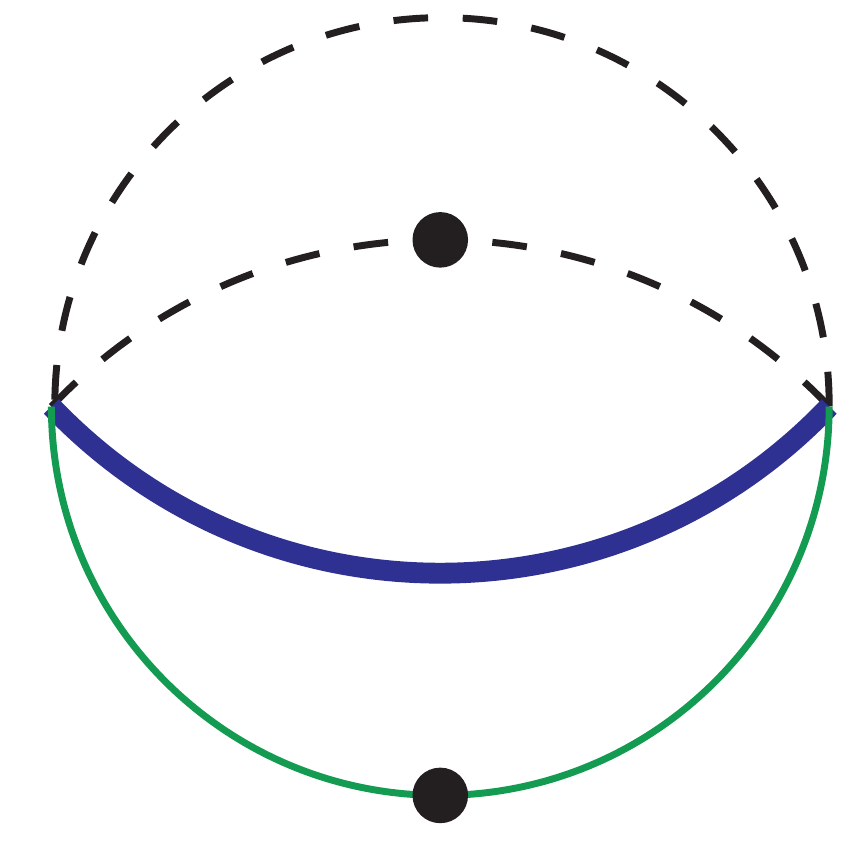}}\qquad
\subfloat[]{\includegraphics[width =2 cm]{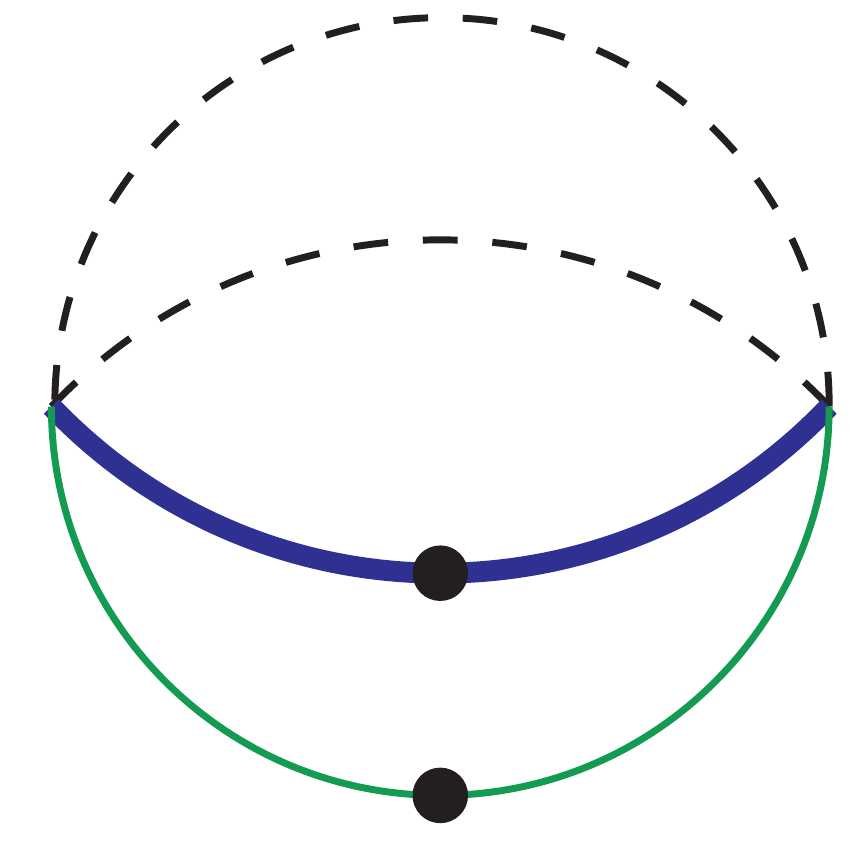}}\\
\subfloat[]{\includegraphics[width =2 cm]{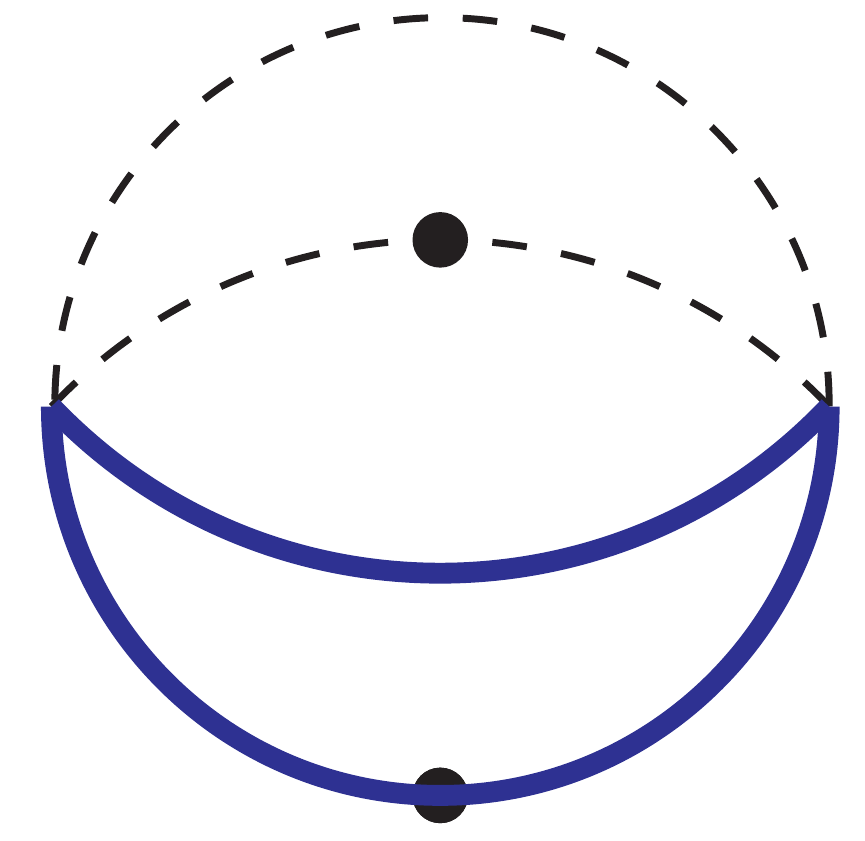}}\qquad
\subfloat[]{\includegraphics[width =2 cm]{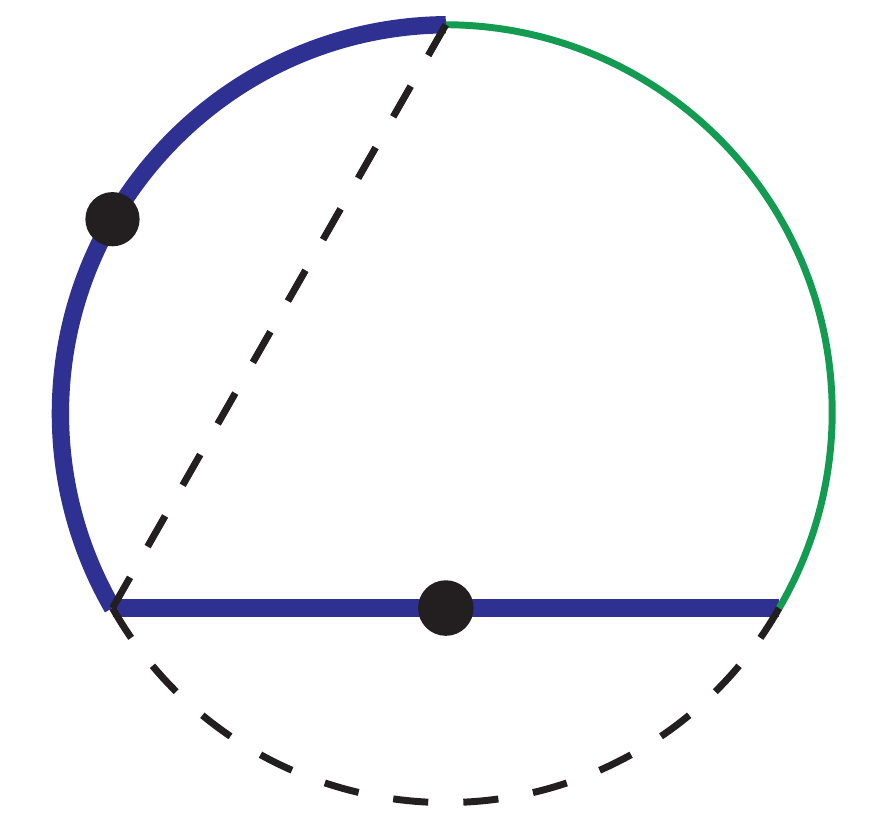}}\qquad
\subfloat[]{\includegraphics[width =2 cm]{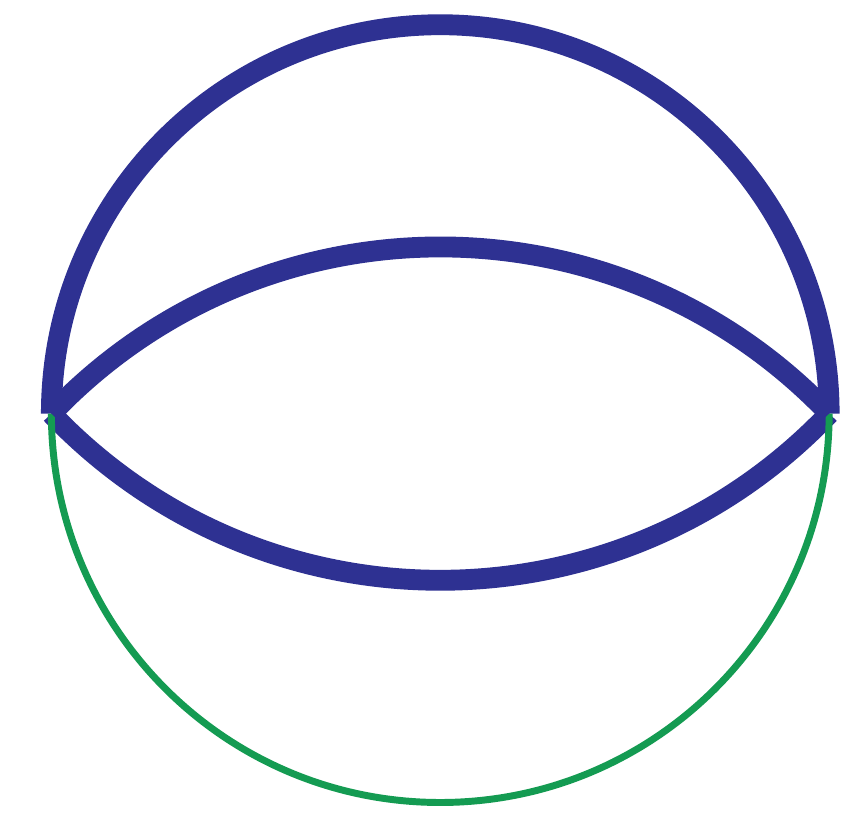}}\qquad
\subfloat[]{\includegraphics[width =2 cm]{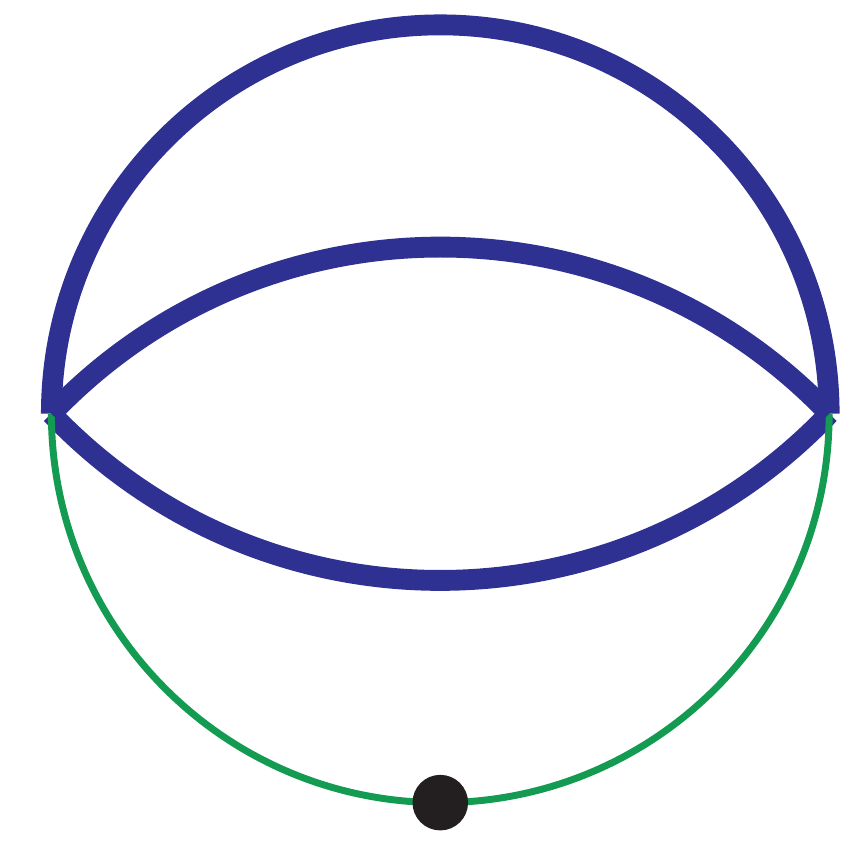}}\qquad
\subfloat[]{\includegraphics[width =2 cm]{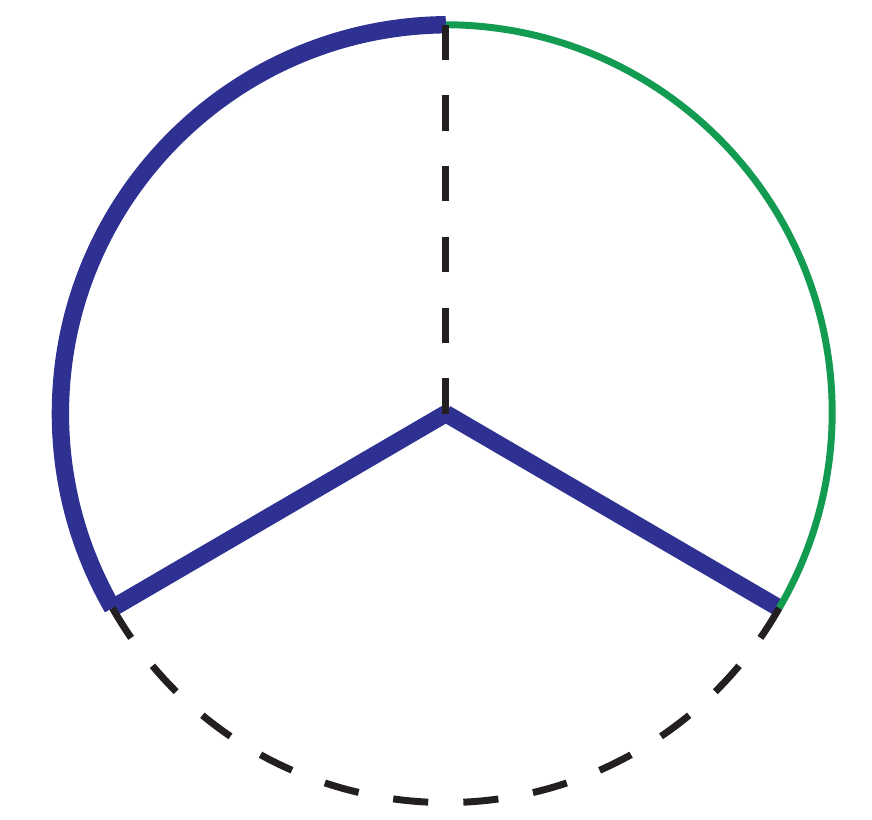}}
\caption{Master integrals for the family of integrals defined in eq.~\eqref{Topo1}. Dots denote squared propagators,
blue (thick) lines denote propagators with mass $m_1^2$,
green (thin) lines denote propagators with mass $m_2^2$, and
dashed lines denote massless propagators.}
\label{fig:allTopo}
\end{figure}

Using Integration-By-Parts (IBP) identities~\cite{Chetyrkin:1981qh,Tkachov:1981wb}, it is possible to reduce all three-loop integrals that contribute to $\delta^{(2)}$ to the  computation of a small set of master integrals~\cite{Grigo:2012ji}. All but six
of the master integrals were evaluated analytically in terms of multiple polylogarithms. The remaining integrals were shown to involve
elliptic functions. They can be embedded in two six-propagator topologies which we call $A$ and $B$, and which only differ by exchanging the roles of $m_1$ and $m_2$ (see fig.~\ref{fig:allTopo}).
Topology $A$ can be written as
\be
\label{Topo1}
\mathcal{J}_{a_1a_2a_3a_4a_5a_6} = \int 
\mathcal{D} k_1 \mathcal{D} k_2 \mathcal{D} k_3 
\frac{1}{D_1^{a_1}D_2^{a_2}D_3^{a_3}D_4^{a_4}D_5^{a_5}D_6^{a_6}}\,,
\ee
where the propagators are defined as
\begin{align}\begin{split}
	D_1=m_1^2-k_1^2\,,\qquad &D_2=m_2^2- k_2^2\,,\qquad D_3=-k_3^2\,,\qquad D_4=m_1^2-(k_1 - k_3)^2\,,\\
	&D_5=-(k_2 - k_1)^2\,,\qquad D_6=m_1^2-(k_3 - k_2)^2\,.
\end{split}\end{align}
We work in dimensional regularisation with $d=4-2\epsilon$, and choose the measure
\begin{equation}
	\mathcal{D} k_l=e^{\gamma_E\epsilon}\frac{d^dk_l}{i\pi^{d/2}}\,.
\end{equation}

Topology $A$ has ten master integrals, whose graphs we list in fig.~\ref{fig:allTopo}. It is convenient  to introduce a variable $t$ defined as\footnote{Here we depart from the conventions of ref.~\cite{Grigo:2012ji} where the results are written
in terms of $x=m_2/m_1$.}
\be
t = \frac{m_2^2}{m_1^2}\,.
\ee 
For concreteness, we will choose $m_2^2<m_1^2$, i.e., $0<t<1$. The case $t>1$ can be obtained by analytic continuation, and we will discuss this at the end of the paper. Note that integrals from Topology $B$ can be identified with integrals from Topology $A$ with $t>1$. Hence, it is sufficient to compute Topology $A$ and to understand the analytic continuation to all positive values of $t$. We therefore only focus on Topology $A$ with $0<t<1$ in the first sections of this paper, and we only comment on Topology $B$ when we discuss the analytic continuation of Topology $A$ to $t>1$. 
For the physical $\rho$ parameter in the SM, we will be interested
in topologies $A$ and $B$ evaluated at $t_\text{phys} = m^2_{\text{b}}/m^2_{\text{t}}$, with
\begin{equation}\label{eq:tPhys}
t_\text{phys} = m^2_{\text{b}}/m^2_{\text{t}}\sim5\cdot10^{-4}\,.
\end{equation}

It is convenient to work with the following basis of master integrals:
\begin{equation}\begin{split}\label{eq:masters}
 f_1(t) &\, =  \epsilon^3 (m_1^2)^{3\epsilon} \mathcal{J}_{0,2,0,2,0,2}\,, \\
f_2(t)  &\,=  \epsilon^3 (m_1^2)^{3\epsilon} \mathcal{J}_{2,0,0,2,0,2}  \,,\\
f_3(t) &\,=\epsilon^3\left(\epsilon-1\right)(m_1^2)^{3\epsilon}\mathcal{J}_{0,2,1,2,0,1}\,,\\
f_4(t)&\,=\epsilon^3\left(\epsilon-1\right)(m_1^2)^{3\epsilon}
\mathcal{J}_{0, 2, 2, 1, 1, 0}\,,\\
f_5(t)  &\,=\epsilon^3\left(\epsilon-1\right)(m_1^2)^{3\epsilon}
\mathcal{J}_{0, 2, 1, 2, 1, 0}\,,\\
f_6(t)  &\,=  \epsilon^3\left(\epsilon - 1\right)(m_1^2)^{3\epsilon}
\mathcal{J}_{2, 0, 2, 0, 1, 1} \,, \\
f_7(t)  &\,=  \epsilon^3\left(\epsilon - 1\right)
(m_1^2)^{3\epsilon} m_2^2\mathcal{J}_{2, 1, 1, 0, 1, 2}  \,,\\
f_8(t) &\, = (m_1^2)^{-2+3\epsilon} \mathcal{J}_{1, 1, 0, 1, 0, 1}  \,,\\
f_9(t) &\, = (m_1^2)^{-1+3\epsilon} \mathcal{J}_{1, 2, 0, 1, 0, 1}\,, \\
f_{10}(t)  &\,=  \epsilon^4 (1-\epsilon)(1-2\epsilon)(m_1^2)^{3\epsilon}
\mathcal{J}_{1, 1, 1, 1, 1, 1}\,.
\end{split}\end{equation}
 We have normalised all master integrals to be dimensionless, i.e., the functions $f_i$ 
only depend on $t$.

An efficient way to compute the master integrals is to use differential equations~\cite{Kotikov:1990kg,Kotikov:1991hm,Kotikov:1991pm,Gehrmann:2000zt,Remiddi:1997ny}. 
The master integrals $f_1$ through $f_7$ 
satisfy a system of differential equations in so-called canonical form~\cite{Henn:2013pwa}:
 \begin{equation}\label{f1f7DE}
 \partial_t f_a = \epsilon \,\frac{(A_0)_{ak}}{t} f_k + \epsilon \frac{(A_1)_{ak} }{t - 1}f_k\,, \qquad 1\le a,k\le 7\,.
 \end{equation}
 $A_0$ and $A_1$ are matrices of integer numbers, 
 which we give explicitly in Appendix~\ref{app:DEQ}.
 The canonical form of the differential equation in eq.~\eqref{f1f7DE}
 makes it manifest that the functions $(f_1,\ldots,f_7)$ can be expressed in terms of a
 well-studied class of special functions 
called
 multiple polylogarithms (MPLs)~\cite{Goncharov2001,goncharov2011multiple}:
 \be\label{eq:mpls}
G(a_1,\cdots,a_n;x) = \int_0^x \frac{du}{u-a_1}G(a_2,\cdots,a_n;u)\,, 
\;\;\; G(;x)=1, \;\;\; a_n\neq 0\,.
\ee
In the case where $a_n=0$, the naive recursive definition in eq.~\eqref{eq:mpls} is divergent, and we define instead,
\beq
G(\underbrace{0,\ldots,0}_{n\textrm{ times}};x) = \frac{1}{n!}\,\log^n x\,.
\eeq
Since eq.~\eqref{f1f7DE} only has singularities at $t\in\{0,1,\infty\}$, we only need to consider the case where $a_i\in\{0,1\}$ in eq.~\eqref{eq:mpls}, which defines a class of functions known as harmonic polylogarithms~\cite{Remiddi:1999ew}. Solving eq.~\eqref{f1f7DE} in terms of MPLs is standard, and the analytic results can be found in ref.~\cite{Grigo:2012ji}. We will therefore not discuss these integrals any further, and we only mention that
we find complete agreement with the results of ref.~\cite{Grigo:2012ji}.

Let us now turn to the remaining master integrals $f_8$, $f_9$ and $f_{10}$.
It was already observed in ref.~\cite{Grigo:2012ji}
that these integrals are not expressible in terms of MPLs.
This can be seen from the perspective of differential equations. Before we discuss the form of the differential 
equations, it is convenient to change the basis of master integrals. 
More precisely, we consider the integrals $f_8$ and $f_9$ in $d=2-2\epsilon$ dimensions where they are finite.
We denote the corresponding master integrals in $d=2$ dimensions by $f_8^{(2)}$ and $f_9^{(2)}$. 
The results in $d=4-2\eps$ dimensions can be recovered from dimensional recurrence relations~\cite{Tarasov:1996br,Lee:2009dh},
which we give explicitly in Appendix~\ref{app:DRR}.
Here we only mention that the poles in $\eps$ of $f_8$ and $f_9$ only involve MPLs, and the two-dimensional integrals 
$f_8^{(2)}$ and $f_9^{(2)}$ enter for the first time in the coefficient of $\eps^0$. It is therefore sufficient 
to compute $f_8^{(2)}$ and $f_9^{(2)}$ (where we set $\eps=0$) to obtain their four-dimensional analogues, which leads to simplifications in our computations. 

Let us now have a closer look at the differential equations satisfied by $f_8^{(2)}$ and $f_9^{(2)}$. 
The differential equations remain coupled in the limit $\eps=0$. The corresponding homogeneous $2\times2$ system 
has the form
\begin{align}\begin{split}\label{eq:diffEll}
\partial_t f_{8,h}^{(2)}  & = - f_{9,h}^{(2)}\,,  \\
\partial_t f_{9,h}^{(2)}  & = \left(\frac{1}{12(t - 9)} + 
\frac{1}{4(t - 1)} - \frac{1}{3 \, t}\right)f_{8,h}^{(2)} -
 \left(\frac{1}{t - 9} + \frac{1}{t - 1} + \frac{1}{t}\right)f_{9,h}^{(2)}\,.
\end{split}\end{align}
It is convenient to transform this system into a second-order equation satisfied by $f_{8,h}^{(2)}$,
\be
\mathcal{D}_t^2f_{8,h}^{(2)} = 0\,,
\ee
with
\be
\label{Dt2}
\mathcal{D}_t^2 = \partial_t^2 + \left(\frac{1}{t - 9} + \frac{1}{t - 1} + 
\frac{1}{t}\right)\partial_t + \left(\frac{1}{12(t-9)} + \frac{1}{4(t-1)} - \frac{1}{3t}\right)\,.
\ee
Remarkably, this differential operator is the same one as in the case of  the sunrise integral with 
a massive external leg and three massive propagators of equal mass~\cite{Laporta:2004rb}.
As a consequence, any element in the kernel of $\mathcal{D}^2_t$ can be written as a
linear combination of the following two functions~\cite{Laporta:2004rb}:
\be
\label{periods}
\Psi_1(t) = \frac{8}{\sqrt{3 + 8\sqrt{t} + 6t - t^2}} \K(\lambda(t)), \;\;\; \Psi_2(t) = \frac{16 i}{\sqrt{3 + 8\sqrt{t} + 6t - t^2}} \K(1 - \lambda(t)),
\ee 
where $\lambda(t)$ is given by
\be\label{eq:lambda_def}
\lambda(t) = \frac{16 \sqrt{t}}{3 + 8\sqrt{t} + 6t - t^2},
\ee
and $\K$ is the complete elliptic integral of the first kind,
\begin{equation}\label{eq:ellK}
\K(\lambda) = \int_0^1\frac{dt}{\sqrt{(1-t^2)(1-\lambda t^2)}}\,.
\end{equation}
The functions in eq.~\eqref{periods} are respectively real and purely imaginary for $0<t<1$. They can be analytically continued outside of this range~\cite{Remiddi:2016gno,Bogner:2017vim}. We will return to the analytic continuation in Section~\ref{Sec7}.

Equation~\eqref{periods} is sufficient to construct the general solution to the homogeneous system in eq.~\eqref{eq:diffEll}. The Wronskian matrix of the system is
\be\label{eq:pMat}
\mathcal{W}(t) = \left(\begin{array}{cc} \Psi_1(t) & \Psi_2(t) \\ \Phi_1(t) & \Phi_2(t) \end{array}\right) \equiv  \left(\begin{array}{cc} \Psi_1(t) & \Psi_2(t) \\ \partial_t\Psi_1(t) & \partial_t\Psi_2(t) \end{array}\right)\,.
\ee
The general solution to the homogeneous equation then takes the form
\beq
\left(\begin{array}{c} f_{8,h}^{(2)}  \\ f_{9,h}^{(2)} \end{array}\right) = \mathcal{W}(t) \left(\begin{array}{c} c_1  \\ c_2 \end{array}\right)\,,\qquad c_i\in\mathbb{C}\,.
\eeq

While the solution of the homogeneous system in eq.~\eqref{eq:diffEll} is well known, 
solving the inhomogeneous system satisfied by $f_8^{(2)}$, $f_9^{(2)}$ and $f_{10}$ is not trivial. 
Here, we will compute $f_8^{(2)}$ from its Feynman parametric representation. 
From their definition in eq.~\eqref{eq:masters}, we see that $f_9^{(2)}$ is determined by the derivative of $f_8^{(2)}$
with respect to $t$.
Once $f_8^{(2)}$ is known in an appropriate form, we can thus compute $f_9^{(2)}$ by differentiating $f_8^{(2)}$. 
Finally, we can compute $f_{10}$ from the differential equation that it satisfies, which reads
\begin{align}\begin{split}\label{EqM10}
&\partial_t f_{10}=\epsilon\frac{2}{1-t}f_{10}+
\epsilon^4\,\Bigg[\frac{2}{t}\,\left(\zeta_3- \text{Li}_3(1-t)\right)-
\frac{1}{3}\left(\frac{4}{1-t}+\frac{1}{2}\right)\,f_8^{(2)}\\
&+\frac{1}{1-t}\,\left(2 \text{Li}_3(t)- \text{Li}_2(t) \log t-
\frac{\pi ^2}{6}\, \log t+\frac{10}{3}\, \zeta_3\right)\Bigg]
+\mathcal{O}(\epsilon^5)\,,
\end{split}\end{align}
where $\textrm{Li}_n(t)$ denote the classical polylogarithms,
\beq
\textrm{Li}_n(t) = -G(\underbrace{0,\ldots,0}_{n-1},1;t)\,,\qquad \zeta_n = \textrm{Li}_n(1)\,.
\eeq
We see that the homogeneous part of the differential equation in eq.~\eqref{EqM10} is in canonical form. Nevertheless,
$f_{10}$ cannot be expressed in terms of MPLs only, because it contains $f_8^{(2)}$ in the inhomogeneous term. 
We can thus determine $f_{10}$ once $f_8^{(2)}$ is known analytically. Note that $f_9^{(2)}$ does not contribute to the differential equation for $f_{10}$ through $\mathcal{O}(\eps^4)$, which is the order through which $f_{10}$ enters the three-loop
corrections to the $\rho$ parameter.

Before we discuss the computation of $f_8^{(2)}$, $f_9^{(2)}$ and $f_{10}$ in the next section, let us mention that the system of differential equations satisfied by $f_8^{(2)}$ and $f_9^{(2)}$
was also analysed in refs.~\cite{Ablinger:2017bjx,Blumlein:2018aeq}. There the corresponding second-order differential operator was not directly in a form which matches the one for the sunrise graph in eq.~\eqref{Dt2}. As a consequence, the homogeneous solutions 
of refs.~\cite{Ablinger:2017bjx,Blumlein:2018aeq} (i.e., the equivalent of our $f_{8,h}^{(2)}$ and $f_{9,h}^{(2)}$) 
take a different form, which cannot easily be recognised as being related to the sunrise graph.


\section{eMPLs on the torus and on the elliptic curve}
\label{sec:3}

As mentioned in the previous section, our strategy for the computation of $f_8^{(2)}$, $f_9^{(2)}$ and $f_{10}$
relies on obtaining an analytic solution for $f_8^{(2)}$ from its Feynman parametric representation.
Before we discuss this in detail, we review in this section some of the mathematical background needed
to perform all the integrals analytically. We keep the review to a strict minimum, and we refer to the literature for 
a more detailed discussion (see,~e.g.,~refs.~\cite{Broedel:2017kkb,Broedel:2018qkq,Broedel:2018iwv} and references therein).

\subsection{Elliptic curves and elliptic functions}
\label{sec:ell&Torus}

Let $P_n(x)$ be a polynomial of degree $n$. An elliptic curve can be
defined (loosely) by the equation $y^2=P_n(x)$ for $n=3,4$. In this paper we are naturally led to an elliptic curve defined by a cubic
polynomial, so we will focus our discussion on the polynomial equation
\be\label{eq:ellCurve}
y^2 = P_3(x) = (x-a_1)(x-a_2)(x-a_3)\,.
\ee
We call $a_1$, $a_2$ and $a_3$ the 
\emph{branch points} of the elliptic curve. 
Seen as points in $\mathbb{CP}^2$ and using homogeneous coordinates, the elliptic curve is given by the points $[x, y,1]$
that satisfy eq.~\eqref{eq:ellCurve}, together with the point $[0,1,0]$ on the 
infinity line. 
It is important to establish our conventions for the branches of the square
root. Throughout this paper we follow ref.~\cite{Broedel:2017kkb}: if the branch points
are real and ordered as $a_1<a_2<a_3$, then 
\be
y = \sqrt{P_3(x)} = \sqrt{\vert P_3(x)\vert}\times \left\{ \begin{array}{l}
-i, \; x \leq a_1 \,,\\
1, \; a_1 < x \leq a_2\,, \\
i, \; a_2 < x \leq a_3\,, \\
-1, \; a_3 < x\,.
\end{array}
\right. 
\ee

Some ubiquitous
quantities that appear in the study of elliptic curves are the 
\emph{periods} $\omega_i$,
\begin{equation}
\omega_1  =  2 c_3 \int_{a_1}^{a_2}\frac{dx}{y} = 2 \K(\lambda)\,, \qquad
\omega_2  =  2c_3 \int_{a_3}^{a_2}\frac{dx}{y} = 2 i \K(1-\lambda)\,,
\end{equation}
with $\K(\lambda)$ defined in eq.~\eqref{eq:ellK}, and the \emph{quasi-periods} $\eta_i$,
\begin{align}\begin{split}\label{eq:quasiPeriods}
\eta_1 &= \frac{1}{4} \int_{a_1}^{a_2}\frac{dx}{c_3y}\left(\frac{s_1(\vec a)}{3}-x\right)
=\E(\lambda)-\frac{2-\lambda}{3}\K(\lambda)\,,\\
\eta_2 &= \frac{1}{4} \int_{a_3}^{a_2}\frac{dx}{c_3y}\left(\frac{s_1(\vec a)}{3}-x\right)
=-i\,\E(1-\lambda)+\frac{1+\lambda}{3}\K(1-\lambda)\,,
\end{split}\end{align}
where 
\be
c_3 = \frac{\sqrt{a_{31}}}{2}\,, \qquad a_{ij} = a_i - a_j\,, 
\qquad \lambda = \frac{a_{12}}{a_{13}}\,,\qquad
s_1(\vec a)=a_1+a_2+a_3\,.
\ee
$\E(\lambda)$ denotes the complete elliptic integral of the second kind,
\begin{equation}\label{eq:ellE}
\E(\lambda) = \int_0^1\sqrt{\frac{1-\lambda t^2}{1-t^2}}dt\,.
\end{equation}
The periods and quasi-periods are related by the Legendre relation:
\begin{equation}
	\omega_1\eta_2-\omega_2\eta_1=-i\pi\,.
\end{equation}

Every elliptic curve is isomorphic to a complex torus. More precisely, if we define
\begin{equation}\label{eq:tau}
	\tau = \frac{\omega_2}{\omega_1}\,,
\end{equation}
then the points in $\mathbb{CP}^2$ that satisfy 
eq.~\eqref{eq:ellCurve} are isomorphic to the 
quotient $\mathbb{C}/\Lambda_{\tau}$ where the two-dimensional lattice~$\Lambda_{\tau}$ is defined as
\begin{equation}
	\Lambda_{\tau}=\mathbb{Z}+\mathbb{Z}\,\tau
	=\left\{m+n\,\tau\,|\,m,n\in\mathbb{Z}\right\}\,.
\end{equation}
Note that $\tau$ can always be chosen to lie in the complex upper half-plane $\mathbb{H}=\{\tau\in\mathbb{C}:\textrm{Im }\tau>0\}$.
We can construct a map from the torus to the elliptic curve with a function
\begin{equation}\label{eq:mu_map}
\mu\left(\cdot,\vec{a}\right): \mathbb{C}/\Lambda_{\tau} \rightarrow \mathbb{C}\,, 
\end{equation}
which satisfies the differential equation 
$\left(c_3\, \mu'\left(z,\vec{a}\right)\right)^2 = 
P_3\left(\mu\left(z,\vec{a}\right)\right)$. A point $z$ on the torus
is then mapped to the point 
$[\mu\left(z,\vec{a}\right), c_3\,\mu'\left(z,\vec{a}\right),1]$ 
on the elliptic curve. The precise form of $\mu\left(z,\vec{a}\right)$
is not relevant for this paper and we refer the reader to ref.~\cite{Broedel:2017kkb} for a more explicit definition.
We can also define a map which assigns to a point $[x_0,y_0,1]$ on the elliptic curve defined by $y^2=P_3(x)$ a point  $z_{x_0}
\in\mathbb{C}/\Lambda_{\tau}$, 
\begin{equation}\label{eq:abel}
z_{x_0} \equiv \frac{c_3}{\omega_1} \int_{\infty}^{x_0} \frac{dx}{y} \;\;\; \operatorname{mod} \; \Lambda_{\tau}\,.
\end{equation}

\subsection{Elliptic multiple polylogarithms}\label{sec:torus}

A natural class of functions to consider when working with Feynman integrals 
are \emph{elliptic multiple polylogarithms} (eMPLs). Loosely speaking, 
eMPLs can be thought of as a class of iterated integrals that generalises
the complete elliptic integrals of eqs.~\eqref{eq:ellK} and \eqref{eq:ellE},
in the same way that the MPLs in eq.~\eqref{eq:mpls} generalise the logarithm.

Since we have two ways of describing an elliptic curve---as a torus $\mathbb{C}/\Lambda_{\tau}$ or as a set of points in 
$\mathbb{CP}^2$---there are also two equivalent ways of defining eMPLs. We start by defining them in terms of iterated integrals along a path on the torus as~\cite{BrownMEP,Broedel_2015}
\be
\label{IItorus}
\tilde{\Gamma}\left(\begin{scriptsize}\begin{array}{lll} n_1  & \cdots & n_k \\ z_1 & \cdots & z_k \end{array}\end{scriptsize};z,\tau \right) = \int_0^z dz' g^{(n_1)}(z'-z_1,\tau)\tilde{\Gamma}\left(\begin{scriptsize}\begin{array}{lll} n_2  & \cdots & n_k \\ z_2 & \cdots & z_k \end{array}\end{scriptsize};z',\tau \right)\,.
\ee
The integer $k$ is called the \emph{length}, $n_i \in \mathbb{N^*}$ and $\sum_in_i$ is the \emph{weight}.
The integration kernels $g^{(n_i)}(z,\tau)$ are defined through the Eisenstein-Kronecker series,
\be
\label{EKs}
F(z,\tau,\alpha) = \frac{1}{\alpha} \sum_{n \geq 0} g^{(n)}(z,\tau) \alpha^n = 
\frac{\theta'_1(0,\tau)\theta_1(z + \alpha,\tau)}{\theta_1(z,\tau)\theta_1(\alpha,\tau)}\,,
\ee
where $\theta_1$ is the Jacobi theta function and $\theta'_1$ is its derivative with respect to the first argument.

Unlike what happens for the MPLs in eq.~\eqref{eq:mpls}, there is an infinite number of kernels one must
consider for eMPLs. While all these kernels are necessary for the study of this class
of functions, in any particular calculation of a Feynman integral only a limited set will be relevant. 
As we will see below, only $g^{(1)}(z,\tau)$ is relevant in our case. From now on we focus on
this kernel and refer the reader to ref.~\cite{BrownMEP} for a more complete exposition. 
First, we note that $g^{(1)}(z,\tau)$ has a simple pole with unit residue at
each point of $\Lambda_{\tau}$. Iterated
integrals over this kernel will thus only have logarithmic singularities. In particular, some of the integrals in
eq.~\eqref{IItorus} may be divergent and require regularisation. We refer the reader to ref.~\cite{Broedel_2015} for details
on how this procedure can be implemented.
Second, $g^{(1)}(z,\tau)$
is odd under $z\to-z$, i.e., $g^{(1)}(-z,\tau)=-g^{(1)}(z,\tau)$. 
Moreover $g^{(1)}(z,\tau)$
is not periodic under translations in both directions of the lattice $\Lambda_{\tau}$:
\begin{equation}
	g^{(1)}(z+1,\tau)=g^{(1)}(z,\tau)\,,\qquad
	g^{(1)}(z+\tau,\tau)=g^{(1)}(z,\tau)-2\pi i \,.
\end{equation}
Finally, it is possible to write a closed form for the total differential of an eMPL.
The total differential for $\tilde{\Gamma}\left(\vec{A};z,\tau\right)$, 
where $\vec{A} = \left(A_1 \cdots A_k\right) = {\scriptsize\left(\begin{array}{ccc}
n_1 & \cdots & n_k \\ z_1 & \cdots & z_k
\end{array}\right)}$, is given by \cite{Broedel:2018iwv}
\bea
\label{dG}
&& d\tilde{\Gamma}\left(A_1\cdots A_k;z,\tau\right) = \sum_{p = 1}^{k-1}(-1)^{n_{p+1}}\tilde{\Gamma}{\scriptsize\left(A_1\cdots A_{p-1} \begin{array}{c}0 \\ 0\end{array}A_{p+2}\cdots A_k; z, \tau\right)} \omega_{p,p+1}^{(n_p + n_{p+1})} \nn \\
&& + \sum_{p = 1}^k\sum_{r = 0}^{n_p+1}\left[ {\scriptsize\left(\begin{array}{c} n_{p-1} + r -1 \\ n_{p-1} - 1\end{array}\right)}\tilde{\Gamma}{\scriptsize \left(A_1 \cdots A_{p-1}^{[r]} \hat{A}_p A_{p+1} \cdots A_k; z , \tau\right)}\omega_{p,p-1}^{(n_p - r)} \right. \nn \\
&& \left. - {\scriptsize\left(\begin{array}{c} n_{p+1} + r -1 \\ n_{p+1} -1 \end{array}\right)}\tilde{\Gamma}{\scriptsize \left(A_1 \cdots A_{p-1} \hat{A}_p A_{p+1}^{[r]}\cdots A_k;z,\tau\right)}\omega_{p,p+1}^{(n_p-r)}\right]\,,
\eea
with $A_p^{[r]} \equiv {\scriptsize\left(\begin{array}{c} n_p + r \\ z_p \end{array} \right)}$. The hat means that the corresponding argument $\hat{A}_p$ is removed, and we use the conventions $\left(z_0,z_{k+1}\right) = (z,0)$, $(n_0,n_{k+1}) = (0,0)$.  
The  $\omega_{ij}^{(n)}$ are differential one-forms given by
\bea
\label{omegaF}
\omega_{ij}^{(n)} & = & (dz_j - dz_i)g^{(n)}(z_j - z_i, \tau) + \frac{n d\tau}{2\pi i}g^{(n+1)}(z_j - z_i,\tau), \;\;\; \operatorname{with} \; n \geq 0\,, \nn \\
\omega_{ij}^{(-1)} & = & - \frac{d\tau}{2\pi i}\,.
\eea

Since we can equivalently represent an elliptic curve as the zero set of some polynomial equation in $\mathbb{CP}^2$, 
see eq.~\eqref{eq:ellCurve},
there is an alternative definition of eMPLs which uses directly the coordinates $(x,y)$ instead of the coordinate $z$ on the torus. 
This representation is important in the context of our current discussion because, in applications to Feynman integral calculations,
elliptic curves often arise via explicit polynomial equations.
The change of variables from $z$ to $x$ is given by eq.~\eqref{eq:mu_map}. Inserting it into the definition
of eMPLs in eq.~\eqref{IItorus} we arrive at the following class of iterated integrals~\cite{Broedel:2017kkb,Broedel:2018qkq}
\begin{equation}\label{cE3}
\mathcal{E}_3\left(\begin{scriptsize}\begin{array}{lll} n_1 \, \cdots \, n_k \\ c_1 \, \cdots 
\, c_k \end{array}\end{scriptsize};x, \vec{a}\right) 
= \int_0^x dt \, \psi_{n_1}(c_1,t,\vec{a})\, \mathcal{E}_3\left(\begin{scriptsize}\begin{array}{lll} n_2 \, 
\cdots \, n_k \\ c_2 \, \cdots \, c_k \end{array}\end{scriptsize};t, \vec{a}\right)\,, 
\qquad \mathcal{E}_3(;x,\vec{a}) = 1\,,
\end{equation}
where the integration kernels are related to the kernels in eq.~\eqref{IItorus} via
\begin{align}
	dx\,\psi_{\pm n}(c,x,\vec{a})
	=dz_x\left(
	g^{(n)}(z_x-z_c,\tau)\pm g^{(n)}(z_x+z_c,\tau)
	-2\delta_{\pm n,1}g^{(1)}(z_x,\tau)
	\right)\,.
\end{align}
Here, $z_x$ is the image of $[x,y,1]$ on the torus, see eq.~\eqref{eq:abel}. The length and the weight of the eMPL 
in eq.~\eqref{cE3} are respectively defined as $k$ and $\sum_{i=1}^k|n_i|$.
Restricting ourselves
to the kernels that will be relevant in this paper, we have
\begin{align}\begin{split}\label{eq:kernelsCE}
	\psi_0(0,x,\vec a)=\frac{c_3}{y \, \omega_1}\,,\quad
	\psi_1(c,x,\vec a)=\frac{1}{x-c}\,,\quad
	\psi_{-1}(c,x,\vec a)=\frac{y_c}{y(x-c)}-\frac{c_3}{2\, y}Z_3(c,\vec a)\,,
\end{split}\end{align}
where
\begin{align}
	Z_3(c,\vec a)=\int_{a_3}^c\frac{dx}{c_3y}\left(\frac{s_1(\vec a)}{3}-x-8c_3^2\frac{\eta_1}{\omega_1}\right)
	=\frac{4g^{(1)}(z_c,\tau)}{\omega_1}\,.
\end{align}
We note that some of the integrals in eq.~\eqref{cE3} may diverge and require regularisation.
We refer to ref.~\cite{Broedel:2018qkq} for a detailed discussion.

Let us make some comments about the iterated integrals defined in this section. First,
the functions defined in eqs.~\eqref{IItorus} and~\eqref{cE3} satisfy the usual properties of iterated integrals.
In particular, they form a shuffle algebra, i.e., any product of these functions evaluated at the same value of the upper integration limit
can be written as a linear combination of the same class of functions.
Second, using eq.~\eqref{eq:kernelsCE} we can easily translate between the eMPLs defined in
eqs.~\eqref{IItorus} and~\eqref{cE3}. Every linear combination of iterated integrals $\tilde{\Gamma}$ 
can be written as a linear combination of $\mathcal{E}_3$ functions, and vice-versa. Finally,
 eMPLs contain MPLs as a subspace. Indeed, the kernel
$\psi_1(c,x,\vec a)$ is precisely the kernel that appears in the definition
of MPLs (c.f. eq.~\eqref{eq:mpls}) and thus
\begin{equation}\label{eq:mToEMPLs}
	\mathcal{E}_3\left(\begin{scriptsize}\begin{array}{lll} 1 \, &\cdots \, &1 \\ c_1 \, &\cdots \, &c_k 
	\end{array}\end{scriptsize};x, 
	\vec{a}\right)=G(c_1,\ldots,c_k;x)\,.
\end{equation}


\section{Evaluating $f^{(2)}_8$ in terms of eMPLs}
\label{sec:f28_eMPL}

In this section we describe the evaluation of $f^{(2)}_8$ in terms of eMPLs directly 
from its Feynman parameter representation. The basic idea is to perform all integrations
in terms of MPLs, except for the last one which can be performed in terms of eMPLs, cf.~refs.~\cite{Hidding:2017jkk,Broedel:2019hyg}.

The Feynman parameter representation for $f^{(2)}_8$ reads
\be
\label{FeynPar}
f_8^{(2)} = \int_{0}^{\infty}\frac{dx_1\,dx_2\,dx_3\,dx_4}{\mathcal{F}}\,\delta\Big(1-\sum_{i\in\Sigma}x_i\Big)\,.
\ee
where $\Sigma$ can be any non-empty subset of $\{1,2,3,4\}$, and
\begin{align}
\mathcal{F} = (x_1 + t x_2 + x_3 + x_4)(x_1 x_2 x_3 + x_1 x_2 x_4 + x_1 x_3 x_4 + x_2 x_3 x_4)\,.
\end{align}
We recall that we can put $\eps=0$ because the integral is finite in $d=2$ dimensions.

We find it convenient to choose $\Sigma = \left\{2\right\}$, which amounts to setting
$x_2=1$ in $\mathcal{F}$. The integrals over $x_1$ and $x_3$ can then be easily performed
in terms of MPLs. After changing variables to 
$\bar{x}_4=\frac{{x_4}}{1+{x_4}}$, we obtain
\begin{align}\begin{split}
\label{LastInt}
f_8^{(2)}  = &\frac{1}{2\sqrt{1-t}} \int_0^1 d\bar{x}_4\frac{1}{y}\left[G_{\pm}\!\left(\chi,0;1\right) -
\log\left(\frac{\bar{x}_4(1-t)+t}{1-\bar{x}_4}\right)\log_\pm\left(1-\frac{1}{\chi}\right)\right.\\
&\left.+G_{\pm}\!\left(\chi,1;1\right)
-G_{\pm}\!\left(\chi,\frac{\bar{x}_4}{\bar{x}_4-1};1\right)
-G_{\pm}\!\left(\chi,\frac{\bar{x}_4(1-t)+t}{\bar{x}_4(2-t)+t-1};1\right)\right]\,,
\end{split}\end{align}
where we used a compact notation which we now explain.
First, we define $y$ as
\begin{align}\label{eq:curveRho}
y^2 =  \left(\bar{x}_4-a_1\right) \left(\bar{x}_4-a_2\right) \left(\bar{x}_4-a_3\right)\,,
\end{align}
with
\begin{equation}
a_1 = \frac{t}{t-1}, \quad a_2 = \frac{1}{8} \left(t+3-\sqrt{(t-1)(t-9)}\right), 
\quad a_3 = \frac{1}{8} \left(t+3+\sqrt{(t-1)(t-9)}\right).
\end{equation}
We note that in the region $0<t<1$ the $a_i$ are real and $a_1<a_2<a_3$.
Then we define $\chi_{\pm}$ as
\begin{equation}
\chi_{\pm}=\frac{2 (t-1) \bar{x}_4^2-3 t \bar{x}_4+t+\bar{x}_4\pm2 \sqrt{1-t} y}{2 (t-1) (\bar{x}_4-1)^2}\,.
\end{equation}
Finally, we use the shorthand
\begin{equation}
f_\pm(\chi)\equiv f(\chi_+)-f(\chi_-)\,.
\end{equation}

While we were able to integrate over three Feynman parameters
without leaving the space of MPLs to reach eq.~\eqref{LastInt}, 
the square-root $y$ appearing in the 
$\bar{x}_4$ integration means
that the last integral will leave this space and should instead be carried in terms of eMPLs. The elliptic curve is defined by eq.~\eqref{eq:curveRho}, and our first step
is to recast the integrand of eq.~\eqref{LastInt} in terms of eMPLs.
We write the integrand of eq.~\eqref{LastInt} as
\begin{equation}
	f^{(2)}_8=\int_{0}^1d\bar{x}_4\,\frac{\Omega(\bar{x}_4;t)}{2\sqrt{1-t}y}\,,
\end{equation}
with
\bea
\label{eq:derOm}
	\frac{\partial \Omega(x;t)}{\partial x} & = &
	\left[
	\frac{1}{\sqrt{1-t}(1-x)y} - \frac{\sqrt{1-t}}{y} - \frac{t}{\sqrt{1-t}xy}
	\right] \nn\\
	&&\times\left[G(0;t)-G(0;x)-G(1;x) 
	+G\left(\frac{t}{t-1};x\right)\right]\,.
\eea
The MPLs appearing in this expression can be written as eMPLs using eq.~\eqref{eq:mToEMPLs}.
Furthermore, since the kernels in eq.~\eqref{eq:derOm} are all of the form of 
the kernels in eq.~\eqref{eq:kernelsCE}, we can write $\Omega(x;t)$ in terms of eMPLs. 
This requires determining the boundary contribution of 
$\Omega(x;t)$ at $x=0$, which is given by
\begin{equation}
\Omega(0;t) =	\frac{\pi^2}{3}+ \log^2t\,.
\end{equation}
Starting from this representation, we can then compute the integral over the last
Feynman parameter $\bar{x}_4$ in terms of eMPLs simply using eq.~\eqref{cE3}, and we find
\bea
\label{f8E3}
	f^{(2)}_8(t) & = & \Psi_1(t)\bigg\{
	\cEf{0 & -1 & 1}{0 & 0 & 0}{1}{\vec{a}}+
	\cEf{0 & -1 & 1}{0 & 0 & 1}{1}{\vec{a}}\\
	&& \left.-\cEf{0 & -1 & 1}{0 & 0 & \frac{t}{t-1}}{1}{\vec{a}}
	-\cEf{0 & -1 & 1}{0 & 1 & 0}{1}{\vec{a}}
	-\cEf{0 & -1 & 1}{0 & 1 & 1}{1}{\vec{a}}
	\right.\nn\\
	&&\left.+\cEf{0 & -1 & 1}{0 & 1 & \frac{t}{t-1}}{1}{\vec{a}} - 2\pi i\,\cEf{0 & 0 & 1}{0 & 0 & 0}{1}{\vec{a}} - 2 \pi i \,\cEf{0 & 0 & 1}{0 & 0 & 1}{1}{\vec{a}} 
	\right.\nn\\
	&& \left. + 2 \pi i \, \cEf{0 & 0 & 1}{0 & 0 & \frac{t}{t-1}}{1}{\vec{a}} + \log t \left[\cEf{0 & -1}{0 & 1}{1}{\vec{a}} - 
	\cEf{0 & -1}{0 & 0}{1}{\vec{a}}
	\right. \right. \nn\\
	&& \left. \left.  + 6\pi i\,\cEf{0 & 0}{0 & 0}{1}{\vec{a}}\right] + \cEf{0}{0}{1}{\vec{a}}\left[\frac{\pi^2}{6} + \frac{\log^2t}{2}\right] \right\}, \nn 
	\eea
where the overall factor $\Psi_1(t)$ is defined in eq.~\eqref{periods}.

Equation~\eqref{f8E3} is one of the main results of this paper and expresses the integral $f^{(2)}_8$ in terms of eMPLs. Let us make some comments about this result. First, we observe that~$f^{(2)}_8$ is proportional to the homogeneous solution $\Psi_1(t)$, which is a period of the elliptic curve defined by the polynomial equation in eq.~\eqref{eq:curveRho}. The period is multiplied by a linear combination of eMPLs of uniform weight two. If we assign weight one to the period~$\Psi_1(t)$~\cite{Broedel:2018qkq}, then $f^{(2)}_8$ has uniform weight three. This is consistent with the fact that $f^{(2)}_8$ can be interpreted as a banana graph in two dimensions with two distinct masses evaluated at zero external momentum. The banana graph in two dimensions is known analytically in the case of three equal masses, and it indeed evaluates to a function of uniform weight three~\cite{Broedel:2019kmn}. Second, we can also express $f^{(2)}_8$ in terms of the eMPLs $\tilde{\Gamma}$. We start by noting that the eMPLs in eq.~\eqref{f8E3} are evaluated at $x\in\{0,1,\frac{t}{t-1}\}$. Under eq.~\eqref{eq:abel} these points are mapped to 
\beq\label{eq:f8_torus_points}
z_0(t) = \frac{1}{3} + \frac{\tau(t)}{2}\,,\qquad z_1(t) = \frac{1}{3}\,,\qquad z_{t/(t-1)}(t) = \frac{\tau(t)}{2}\,,
\eeq
where $\tau$ denotes the ratio of the two periods of the elliptic curve (see eq.~\eqref{periods}),
\beq\label{eq:tau_to_psi}
\tau(t) = \frac{\Psi_2(t)}{\Psi_1(t)}\,.
\eeq
Since there is never any ambiguity, we will in the following not explicitly write the dependence of the $z_i$ and $\tau$ on $t$.
We can then use this to rewrite the integration kernels~\eqref{eq:kernelsCE} in terms of those found in the definition of the $\tilde{\Gamma}$, and we find

\bea
\label{PsiToG}
dx\,\psi_0(0,x,\vec{a}) & = & dz, \nn \\
dx\,\psi_{-1}(0,x,\vec{a}) & = & dz\left[g^{(1)}\left(z -\frac{1}{3} - \frac{\tau}{2},\tau\right) - g^{(1)}\left(z + \frac{1}{3} + \frac{\tau}{2},\tau\right)\right], \nn \\
dx\,\psi_{-1}(1,x,\vec{a}) & = & dz\left[g^{(1)}\left(z - \frac{1}{3},\tau\right) - g^{(1)}\left(z + \frac{1}{3},\tau\right)\right], \nn \\
dx\,\psi_{1}(0,x,\vec{a}) & = & dz\left[g^{(1)}\left(z -\frac{1}{3}-\frac{\tau}{2},\tau\right) + g^{(1)}\left(z + \frac{1}{3}+\frac{\tau}{2} ,\tau\right) - 2 g^{(1)}(z,\tau)\right], \nn \\
dx\,\psi_{1}(1,x,\vec{a}) & = & dz\left[g^{(1)}\left(z - \frac{1}{3},\tau\right) + g^{(1)}\left(z + \frac{1}{3},\tau\right)- 2 g^{(1)}(z,\tau)\right], \nn \\
dx\,\psi_{1}\left(\frac{t}{t-1},x,\vec{a}\right) & = & dz\left[g^{(1)}\left(z - \frac{\tau}{2},\tau\right) - g^{(1)}\left(z + \frac{\tau}{2},\tau\right) - 2 g^{(1)}(z,\tau)\right].
\eea
Using this change of variables in eq.~\eqref{cE3}, we can express all the $\mathcal{E}_3$ functions in eq.~\eqref{f8E3} in terms of $\tilde{\Gamma}$ functions. The procedure is straightforward, but the result is lengthy and not particularly illuminating, so we do not show it here explicitly. We only mention that since all points in eq.~\eqref{eq:f8_torus_points} are rational points of the form $z_i = \frac{r}{6}+\frac{s\,\tau}{6}$, with $r$ and $s$ integers, the $\tilde{\Gamma}$ functions will have a very special form, namely they will all have arguments that are rational points.

Our next goal is to compute the master integrals $f_9^{(2)}$ and $f_{10}$. As explained at the end of Section~\ref{sec:notAndConv}, this can be done by differentiating or integrating $f_8^{(2)}$ with respect to $t$. These operations, however, are not straightforward to carry out on the expressions in eq.~\eqref{f8E3}, because the eMPLs depend on $t$ in a highly non-trivial way. It would be desirably to have a representation of $f_8^{(2)}$ in terms of iterated integrals with a simple dependence on the kinematic variable. While in general such a form may not be easily obtained, the special rational form of the points in eq.~\eqref{eq:f8_torus_points} allows one to find such a representation in this case. This will be reviewed in the next section.


\section{eMPLs and iterated Eisenstein integrals}
\label{Sec5}

In this section we review how eMPLs evaluated at rational points can be expressed in terms of another class of iterated integrals, namely the so-called
iterated Eisenstein integrals. Eisenstein series are a special case of modular forms. In the first part of this section we review modular forms and Eisenstein series, and
in a second part we review the relationship between iterated Eisenstein integrals and eMPLs.

\subsection{Modular forms: a brief introduction}

In Section~\ref{sec:torus} we have seen that that every elliptic curve is isomorphic to a torus $\mathbb{C}/\Lambda_\tau$,
for some value $\tau\in\mathbb{H}$.
Different values of $\tau$, however, do not necessarily
describe different tori. Indeed, we can replace the basis of periods $(\omega_1,\omega_2)$ by an integer linear combination of 
them without changing the lattice they generate. More precisely, let $\tau'\in\mathbb{H}$ be obtained from $\tau$ via a \emph{modular transformation}, defined as
\be\label{eq:modTrans}
\tau\to\tau' = \gamma \cdot \tau \equiv \frac{a\tau + b}{c\tau + d}\,, \qquad
\gamma = \left(\begin{smallmatrix} a & b \\ c & d \end{smallmatrix}\right) \in \SL(2,\mathbb{Z}),
\ee
where
\begin{equation}
	\SL(2,\mathbb{Z})=\Big\{
	\left(\begin{smallmatrix} a & b \\ c & d \end{smallmatrix}\right)
	\big|a,b,c,d\in\mathbb{Z},\,ad-bc=1
	\Big\} \,.
\end{equation}
Then $\tau$ and $\tau'$ define the same lattice, $\Lambda_{\tau} = \Lambda_{\tau'}$, and so they also define the same elliptic curve.
In other words, 
$\tau$ and $\tau'$ define the same elliptic curve if and only if they are related
by a modular transformation. For this reason, modular transformations play a central
role in the study of elliptic curves. 

In applications it is often specific subgroups of $\SL(2,\mathbb{Z})$ that are of interest. 
Particularly important subgroups are the so-called \emph{congruence subgroups of level $N$}, defined by
\beq\begin{split}
\label{CongSub}
\Gamma_0(N)  = \,& \Big\{ \left(\begin{smallmatrix} a & b \\ c & d \end{smallmatrix}\right) 
\in \SL(2,\mathbb{Z})\, \big| \, c = 0 \, \operatorname{mod} \, N\Big\}\,,  \\
\Gamma_1(N)  = \,& \Big\{\left(\begin{smallmatrix} a & b \\ c & d \end{smallmatrix}\right) 
\in \SL(2,\mathbb{Z})\, \big| \, c = 0 \, \textrm{~and~} \, a=d=1 \, \operatorname{mod} \, N\Big\}\,,  \\
\Gamma(N)  = \,& \Big\{ \left(\begin{smallmatrix} a & b \\ c & d \end{smallmatrix}\right) 
\in \SL(2,\mathbb{Z})\, \big| \, b = c = 0 \, \textrm{~and~} \, a=d=1 \, \operatorname{mod} \, N\Big\}\,,
\end{split}\eeq
with $\Gamma(N)\subseteq\Gamma_1(N)\subseteq\Gamma_0(N)\subseteq \SL(2,\mathbb{Z})$.
Let $\Gamma$ be any congruence subgroup of $\SL(2,\mathbb{Z})$. It is clear that 
$\mathbb{H}$ maps onto itself under modular transformations for $\Gamma$.
If we consider the extended upper-half plane 
$\overline{\mathbb{H}}=\mathbb{H}\,\cup\mathbb{Q}\,\cup\left\{i \infty\right\}$, then
$\Gamma$ acts separately on $\mathbb{H}$ and on $\mathbb{Q}\,\cup\left\{i \infty\right\}$.
The equivalence classes of $\mathbb{Q}\,\cup\left\{i \infty\right\}$ are called the \emph{cusps}
of $\Gamma$ (we recall that $\tau$ and $\tau'$ are in the same
equivalence class if $\tau=\gamma\cdot\tau'$ for some $\gamma\in\Gamma$). The equivalence class
that contains $i\infty$ is called the \emph{cusp at infinity}.
As an example, for $\gamma\in \Gamma(1)= \SL(2,\mathbb{Z})$ we find that 
$\gamma\cdot(i\infty)=a/c$ and thus there is a single cusp for $\Gamma(1)$,
the cusp at infinity. For any $N$, the number of cusps of the congruence subgroups of
eq.~\eqref{CongSub} is finite.

Our goal is to construct functions that transform nicely under some congruence subgroup $\Gamma$.
A \emph{modular function for $\Gamma$} is a meromorphic function $f:\overline{\mathbb{H}}\to \mathbb{C}$ that is invariant under $\Gamma$. 
One can show that every modular function has at least one pole. 
If we want to consider holomorphic functions, i.e., functions without poles, then we need to consider more general transformations.
A \emph{modular form of weight $n$ for $\Gamma$} is a function
$f:\overline{\mathbb{H}}\to\mathbb{C}$ that is holomorphic on 
$\mathbb{H}$ and at the cusps of $\Gamma$ such that
\be
\label{ModT}
f\left(\gamma\cdot\tau\right) = (c\tau + d)^n f(\tau)\,,\qquad
\gamma= \left(\begin{smallmatrix} a & b \\ c & d \end{smallmatrix}\right) \in\Gamma\,.
\ee
Let $N$ be the smallest integer such that $\Gamma(N)\subseteq\Gamma$.
Then translations by $N$ are generated by $T_N=\left(\begin{smallmatrix} 1 & N \\ 0 & 1 \end{smallmatrix}\right)$
and so modular forms in $\Gamma$ are $N$-periodic. In particular,
this implies that modular forms of level $N$ admit a Fourier expansion of
the form
\be
\label{Qexp}
f(\tau) = \sum_{m = 0}^{\infty} a_m e^{\frac{2 \pi i m \tau}{N}} = \sum_{m = 0}^{\infty} a_m q_N^m\,,
\ee
where $q = \exp(2 \pi i \tau)$ and $q_N = q^{\frac{1}{N}}$, and we used
the fact that $f(\tau)$ is holomorphic at $\tau\to i\infty$ 
to start the summation at $m=0$. This Fourier expansion is called the 
\emph{$q$-expansion} of  the modular form $f(\tau)$.

Let us denote by $\mathcal{M}_n(\Gamma)$ the vector space generated by all modular forms of weight $n$ for $\Gamma$.
One can show that $\mathcal{M}_n(\Gamma)$ is always finite-dimensional. Moreover, $\mathcal{M}_n(\Gamma)$ admits a direct sum decomposition
\beq
\mathcal{M}_n(\Gamma) = \mathcal{S}_n(\Gamma) \oplus \mathcal{E}_n(\Gamma)\,.
\eeq
Here $\mathcal{S}_n(\Gamma)$ denotes the space of \emph{cusp forms} of weight $n$, i.e., the modular forms of weight $n$ that vanish at all cusps of $\Gamma$.
Its complement $\mathcal{E}_n(\Gamma)$ is the space of Eisenstein series. The Eisenstein series are of particular interest when working with eMPLs, 
and so we discuss them in detail in the remainder of this section.

Of special importance here will be the functions~\cite{Duhr:2019rrs} ($0 \leq r,s < N$),
\begin{align}\begin{split}\label{eq:defAandB}
	\textbf{a}^{(n)}_{N,r,s}(\tau)&=
	\frac{1}{2}\sum_{\substack{(a,b)\in\mathbb{Z}^2\\(a,b)\neq(0,0)}}
	\frac{e^{-2\pi i a r/N}\cos\frac{2\pi s b}{N}}{(a\tau+b)^n}\,,\\
	\textbf{b}^{(n)}_{N,r,s}(\tau)&=
	\frac{1}{2i}\sum_{\substack{(a,b)\in\mathbb{Z}^2\\(a,b)\neq(0,0)}}
	\frac{e^{-2\pi i a r/N}\sin\frac{2\pi s b}{N}}{(a\tau+b)^n}\,.
\end{split}\end{align}
One can show that these functions are always Eisenstein series of weight $n$ for $\Gamma(N)$~\cite{Broedel:2018iwv,Duhr:2019rrs}. Moreover, they form a spanning set of 
$\mathcal{E}_n(\Gamma(N))$, i.e., every element of $\mathcal{E}_n(\Gamma(N))$ can be written as a linear combination of 
the functions in eq.~\eqref{eq:defAandB}. Note that they do not form a basis. The relations between them are however understood~\cite{Broedel:2018iwv,Duhr:2019rrs}.
We also mention that these functions are real whenever $\tau$ is purely imaginary~\cite{Duhr:2019rrs}. Finally, the Eisenstein series in eq.~\eqref{eq:defAandB}
are closely related to the coefficients $g^{(n)}(z,\tau)$ in eq.~\eqref{EKs}. More precisely, we have the relation ($0\le r,s<N$)~\cite{Broedel:2018iwv,Duhr:2019rrs}
\be
\label{gToAB}
g^{(n)}\left(\frac{r}{N} + \frac{s}{N}\tau,\tau\right) = 
-\sum_{k = 0}^{n} \frac{(-2\pi is)^k}{k!\,N^k}\left[\textbf{a}^{(n-k)}_{N,r,s}(\tau)+
i\,\textbf{b}^{(n-k)}_{N,r,s}(\tau)
\right]\,.
\ee
The previous equation shows that there is a connection between Eisenstein series for $\Gamma(N)$ and eMPLs evaluated at rational points of the form
$z = \frac{r}{N} + \frac{s}{N}\tau$. We know from eq.~\eqref{eq:f8_torus_points} that the arguments of the eMPLs that appear in $f_8^{(2)}$ have this form, with $N=6$. 
Hence we expect the integral $f_8^{(2)}$ to be closely connected to Eisenstein series. We review this connection in the next section.

\subsection{Iterated Eisenstein integrals}

Consider a set of modular forms $f_j(\tau)$ of weight $n_j$ for some congruence subgroup of level~$N$. 
We define their iterated integral as~\cite{ManinModular,2014arXiv1407.5167B}
\begin{equation}\label{IIh}
	I(f_1,\ldots,f_k;\tau)=\int_{i\infty}^\tau\frac{d\tau'}{2\pi i}
	f_1(\tau')I(f_2,\ldots,f_k;\tau')\,,
\end{equation}
with $I(;\tau)\equiv1$.
A precise definition of these integrals requires a careful regularisation of divergences
that can appear at the cusp at infinity $\tau'=i\infty$, and we refer to ref.~\cite{2014arXiv1407.5167B} for a detailed discussion.
We define the length of $I(f_1,\ldots,f_k;\tau)$ to be $k$ and the weight is $-k+\sum_{j=1}^kn_j$.
In the case where all the modular forms $f_j(\tau)$ are Eisenstein series, we refer to the integral in eq.~\eqref{IIh}
as an \emph{iterated Eisenstein integral}. 

In ref.~\cite{Broedel:2018iwv} it was shown that whenever an eMPL of weight $n$ is evaluated at rational points of the form
$z = \frac{r}{N} + \frac{s}{N}\tau$, then this eMPL can be expressed as a linear combination of uniform weight $n$ of iterated Eisenstein integrals 
of level $N$. More precisely, we see from eq.~\eqref{dG} that the total differential of a length-$k$ eMPL 
is given by eMPLs of length $k-1$ and one-forms $\omega_{ij}^{(n)}$. If all the arguments of the eMPL are rational,
then we can write $\omega_{ij}^{(n)}$ in terms of Eisenstein series using eq.~\eqref{gToAB}. Hence, if we assume recursively
that the claim is true for eMPLs up to length $k-1$, we see that the total differential of an eMPL of length $k$ evaluated at rational points only involves
iterated Eisenstein integrals and Eisenstein series. We can then integrate back in $\tau$ to obtain the desired representation at length $k$.

This recursive consideration also provides an efficient algorithm to express eMPLs evaluated at rational points in terms of iterated Eisenstein integrals.
The starting point is to write an eMPL evaluated at rational points as the integral of its derivative with respect to~$\tau$,
\begin{equation}
\tilde{\Gamma}\left(A_1\cdots A_k;z,\tau\right)
=\operatorname{Cusp}\left(\tilde{\Gamma}\left(A_1\cdots A_k;z,\tau\right)\right) + \int_{i\infty}^\tau d\tilde{\Gamma}\left(A_1\cdots A_k;z,\tau\right)\,.
\end{equation}
Since the differential lowers the length by one, we can recursively express the integrand in terms of iterated Eisenstein integrals and integrate back using eq.~\eqref{IIh}. 
The integration constant is obtained by studying the behaviour at the cusp at infinity.
For the cases of interest here, it is possible to compute $\operatorname{Cusp}\left(\tilde{\Gamma}\left(A_1\cdots A_k;z,\tau\right)\right)$ by performing a series expansion of the integrands in $\tilde{\Gamma}\left(A_1\cdots A_k;z,\tau\right)$, and integrating only the leading terms (see the example below).
This algorithm is iterative in the length of the eMPLs,
and the starting point is the total differential of an eMPL of length one,
\bea
\label{dG1}
d\, \tilde{\Gamma}{\scriptsize \left(\begin{array}{c} n_1 \\ z_1 \end{array};z,\tau\right)} & = & \sum_{r = 0}^{n_1 + 1}{\scriptsize\left(\begin{array}{c} r-1 \\ -1 \end{array}\right)}\left[\omega_{1,0}^{(n_1 - r)} - \omega_{1,2}^{(n_1 -r)}\right]\,
\eea
where the differential one-forms $\omega_{ij}^{(n_1 - r)}$ can be expressed in terms of Eisenstein series via eq.~\eqref{gToAB} whenever $z_1$ and $z$ are rational points.

To make the discussion more concrete, we illustrate this procedure on the example of one of the eMPLs that appears in the analytic result for $f_8^{(2)}$, namely
$\tilde{\Gamma}{\left(\begin{smallmatrix} 0 & 1 \\ 0 & \frac{2}{3}\end{smallmatrix}; \frac{\tau}{2}, \tau\right)}$.
We start by computing its total differential using eq.~\eqref{dG}. We get:
\begin{align}
\label{GdG}
d \tilde{\Gamma}&{\scriptsize\left(\begin{array}{cc} 0 & 1 \\ 0 & \frac{2}{3}\end{array}; \frac{\tau}{2}, \tau\right)} = 
\sum_{r=0}^{1}\left[
{\scriptsize\left(\begin{array}{c} r-1 \\ -1 \end{array}\right)}\tilde{\Gamma}{\scriptsize \left(\begin{array}{c} 1 \\ \frac{2}{3}\end{array}; \frac{\tau}{2},\tau \right)}\omega^{(-r)}_{1,0} - 
{\scriptsize \left(\begin{array}{c} r \\ 0 \end{array}\right)}\tilde{\Gamma}{\scriptsize \left(\begin{array}{c} 1 + r \\ \frac{2}{3}\end{array}; \frac{\tau}{2},\tau\right)}\omega^{(-r)}_{1,2}
\right]  \\
& + \sum_{r=0}^{2}\left[{\scriptsize\left(\begin{array}{c} r-1 \\ -1 \end{array}\right)}\tilde{\Gamma}{\scriptsize \left(\begin{array}{c} r \\ 0 \end{array}; \frac{\tau}{2},\tau \right)}\omega^{(1-r)}_{2,1} - {\scriptsize \left(\begin{array}{c} r -1 \\ -1 \end{array}\right)}\tilde{\Gamma}{\scriptsize \left(\begin{array}{c} 0 \\ 0 \end{array}; \frac{\tau}{2},\tau\right)}\omega^{(1-r)}_{2,3}\right] 
- \tilde{\Gamma}{\scriptsize\left(\begin{array}{c} 0 \\ 0 \end{array};\frac{\tau}{2},\tau\right)}\omega^{(1)}_{1,2}  \nn \\
& =  \tilde{\Gamma}{\scriptsize \left(\begin{array}{c} 0 \\ 0 \end{array}; \frac{\tau}{2}, \tau\right)}\left[\omega_{2,1}^{(1)} - \omega_{2,3}^{(1)} - \omega_{1,2}^{(1)}\right] + \tilde{\Gamma}{\scriptsize \left(\begin{array}{c} 1 \\ \frac{2}{3} \end{array}; \frac{\tau}{2},\tau\right)}\left[\omega^{(0)}_{1,0} - \omega^{(0)}_{1,2}\right] - \tilde{\Gamma}{\scriptsize\left(\begin{array}{c} 2 \\ \frac{2}{3} \end{array}; \frac{\tau}{2}, \tau\right)}\omega^{(-1)}_{1,2}. \nn 
\end{align}
Iterating this procedure, we find that the total differentials of the eMPLs in the right-hand side of eq.~\eqref{GdG}  read:
\begin{align}\begin{split}
\label{GdGsub}
d \tilde{\Gamma}{\scriptsize\left(\begin{array}{c} 0 \\ 0 \end{array}; \frac{\tau}{2}, \tau\right)}  = & \,\omega^{(0)}_{1,0} - \omega^{(0)}_{1,2}\,,\qquad
d \tilde{\Gamma}{\scriptsize\left(\begin{array}{c} 1 \\ \frac{2}{3} \end{array}; \frac{\tau}{2}, \tau\right)}  = \omega^{(1)}_{1,0} - \omega_{1,2}^{(1)}\,,\\
&d \tilde{\Gamma}{\scriptsize\left(\begin{array}{c} 2 \\ \frac{2}{3} \end{array}; \frac{\tau}{2}, \tau\right)}  =  \omega^{(2)}_{1,0} - \omega^{(2)}_{1,2}\,.
\end{split}\end{align}
The next step is to rewrite the one-forms $\omega^{(n)}_{ij}$ that appear in eqs.~\eqref{GdG} and \eqref{GdGsub} in terms
of the $\textbf{a}^{(n_j)}_{N_j,r_j,s_j}(\tau)$ and $\textbf{b}^{(n_j)}_{N_j,r_j,s_j}(\tau)$. 
Starting from eq.~\eqref{omegaF} and using eq.~\eqref{gToAB}, we find 
\begin{align}\begin{split}
\label{eq:om_ex_1}
\omega^{(1)}_{2,1} & =  -\frac{d\tau}{2 \pi i}\,\textbf{a}^{(2)}_{6,2,0}(\tau)\,,  \\
\omega^{(1)}_{1,0} & =  -\frac{d\tau}{2 \pi i}\,\left[-\frac{\pi^2 }{2} - \textbf{a}^{(2)}_{6,1,0}(\tau) -\textbf{a}^{(2)}_{6,1,3}(\tau)\right]\,,  \\
\omega^{(2)}_{1,0} & =  -\frac{d\tau}{2 \pi i}\,\left[\frac{i\pi^3 d\tau}{2} + i\pi\, \textbf{a}^{(2)}_{6,1,0}(\tau) +i\pi\, \textbf{a}^{(2)}_{6,1,3}(\tau) +2\, \textbf{a}^{(3)}_{6,2,3}(\tau)\right]\,,  \\
\omega^{(2)}_{1,2} & =  -\frac{2d\tau}{3 \pi i}\,\left[\textbf{a}^{(3)}_{6,1,0}(\tau) + \textbf{a}^{(3)}_{6,1,3}(\tau) - \textbf{a}^{(3)}_{6,2,3}(\tau)\right]\,.
\end{split}\end{align}
Inserting eq.~\eqref{eq:om_ex_1} into eq.~\eqref{GdGsub} and integrating back in $\tau$, we find
\bea
\tilde{\Gamma}{\scriptsize\left(\begin{array}{c} 0 \\ 0 \end{array}; \frac{\tau}{2}, \tau\right)} & = & i \, \pi\,I(1;\tau)\,,  \\
\tilde{\Gamma}{\scriptsize\left(\begin{array}{c} 1 \\ \frac{2}{3} \end{array}; \frac{\tau}{2}, \tau\right)} & = & 
\frac{\pi^2}{2}I{ \left(1;\tau\right)} + \IEs{\textbf{a}^{(2)}_{6,1,0}}{\tau} + \IEs{\textbf{a}^{(2)}_{6,1,3}}{\tau} + I{\left(\textbf{a}^{(2)}_{6,2,0};\tau\right)} - G\!\left(e^{-\frac{2 i \pi}{3}};1\right)\,, \nn \\
\tilde{\Gamma}{\scriptsize\left(\begin{array}{c} 2 \\ \frac{2}{3} \end{array}; \frac{\tau}{2}, \tau\right)} & = & 
-\frac{i \, \pi^3}{6}I{\scriptsize\left(1;\tau\right)} - i \, \pi I{\left(\textbf{a}^{(2)}_{6,1,0};\tau\right)} - i \, \pi I{\left(\textbf{a}^{(2)}_{6,1,3};\tau\right)} + \frac{4}{3} I{\left(\textbf{a}^{(3)}_{6,1,0};\tau\right)}  \nn \\
\nn&&+ \frac{4}{3} I{\left(\textbf{a}^{(3)}_{6,1,3};\tau\right)} - \frac{10}{3} I{\left(\textbf{a}^{(3)}_{6,2,3};\tau\right)}\,,
\eea
where we used boundary conditions:
\begin{align}\begin{split}\label{eq:cuspval}
\operatorname{Cusp}\left(\tilde{\Gamma}{\scriptsize\left(\begin{array}{c} 0 \\ 0 \end{array}; \frac{\tau}{2}, \tau\right)}\right) &\,= \operatorname{Cusp}\left(\tilde{\Gamma}{\scriptsize\left(\begin{array}{c} 2 \\ \frac{2}{3} \end{array}; \frac{\tau}{2}, \tau\right)}\right)  =  0\,,\\
\operatorname{Cusp}\left(\tilde{\Gamma}{\scriptsize\left(\begin{array}{c} 1 \\ \frac{2}{3} \end{array}; \frac{\tau}{2}, \tau\right)}\right)
  &\,=   - G\!\left(e^{-\frac{2 i \pi}{3}};1\right)\,.
\end{split}\end{align}
As previously said, the cusp values can be computed from the series expansion of the integration kernels of the eMPLs. 
Let us consider $\tilde{\Gamma}{\scriptsize\left(\begin{array}{c} 1 \\ \frac{2}{3} \end{array}; \frac{\tau}{2}, \tau\right)}$. 
The series expansion of $g^{(1)}\left(z - \frac{2}{3},\tau\right)$ is given by:
\be \label{eq:gsexp}
g^{(1)}\left(z - \frac{2}{3},\tau\right) = \frac{i \pi}{e^{2\pi i (z + \frac{1}{3})} - 1} + i\pi \frac{e^{2\pi i (z + \frac{1}{3})}}{e^{2\pi i (z + \frac{1}{3})} - 1} + \mathcal{O}(q_2)\,.
\ee
Then, performing the change of variable
\be
w =  e^{2\pi i z},
\ee
and integrating eq.~\eqref{eq:gsexp} with respect to $w$ we obtain:
\be\begin{split} \label{eq:Gexp}
\tilde{\Gamma}{\scriptsize\left(\begin{array}{c} 1 \\ \frac{2}{3} \end{array}; \frac{\tau}{2}, \tau\right)} &\,= G\left(e^{-\frac{2 \pi i}{3}};e^{i \pi \tau}\right) - \frac{1}{2}G\left(0; e^{i \pi \tau}\right) - G\left(e^{-\frac{2 \pi i}{3}};1\right) + \mathcal{O}(q_2)\\
&\,=  - \frac{1}{2}\log q_2 - G\left(e^{-\frac{2 \pi i}{3}};1\right) + \mathcal{O}(q_2)\,.
\end{split}\ee
The value of 
$\operatorname{Cusp}\left(\tilde{\Gamma}{\scriptsize\left(\begin{array}{c} 1 \\ \frac{2}{3} \end{array}; 
\frac{\tau}{2}, \tau\right)}\right)$ is defined  as the constant term
in the above equation (i.e., the term independent of $q_2$), which gives the result in eq.~\eqref{eq:cuspval}.

Substituting \eqref{eq:cuspval} into eq.~\eqref{GdGsub}
and using the fact that
\bea
\operatorname{Cusp}\left(\tilde{\Gamma}{\scriptsize\left(\begin{array}{cc} 0 & 1 \\ 0 & \frac{2}{3}\end{array}; \frac{\tau}{2}, \tau\right)}\right) & = & \frac{i}{2\pi}G{\scriptsize\left(0,e^{-\frac{2 i \pi}{3}};1\right)}\,,
\eea
we finally find that
\begin{align}\begin{split}
\tilde{\Gamma}{\scriptsize\left(\begin{array}{cc} 0 & 1 \\ 0 & \frac{2}{3}\end{array}; \frac{\tau}{2}, \tau\right)} 
= & \frac{i}{2\pi}G{\scriptsize\left(0,e^{-\frac{2 i \pi}{3}};1\right)}  - \pi \, i \, G\left(e^{-\frac{2 i \pi}{3}};1\right)I{\scriptsize\left(1;\tau\right)} + \frac{i\, \pi^3}{3}I\left(1,1;\tau\right)  \\
& + i \, \pi \, I\left(1,\textbf{a}^{(2)}_{6,2,0};\tau\right) + \frac{4}{3}I\left(1,\textbf{a}^{(3)}_{6,1,0};\tau\right) + \frac{4}{3}I\left(1,\textbf{a}^{(3)}_{6,1,3};\tau\right) \\
& - \frac{10}{3}I\left(1,\textbf{a}^{(3)}_{6,2,3};\tau\right) + i \, \pi \, I\left(\textbf{a}^{(2)}_{6,2,0},1;\tau\right).
\end{split}\end{align}
Following exactly the same steps as in this example, we can express all the eMPLs that appear in $f_8^{(2)}$ in terms of iterated Eisenstein integrals. The result will be presented in the next section.

We finish this section by noting that the procedure that we just presented to relate eMPLs and iterated Eisenstein integrals
can be reformulated in terms of the coaction on eMPLs \cite{BrownNMP,Broedel:2018iwv}. Indeed, the algorithm
we presented is summarised by the relation
\be
\tilde{\Gamma}\left(\vec{A};z,\tau\right) = m\left[
\left(\operatorname{Cusp}\otimes \int_{i\infty}^\tau d\tau'\right) \Delta \left(\tilde{\Gamma}
\left(\vec{A};z,\tau'\right)\right)\right],
\ee
where $\Delta \left(\tilde{\Gamma}\left(\vec{A};z,\tau\right)\right)$ is the coaction on the eMPLs 
and we defined $m[a\otimes b]\equiv ab$.


\section{The master integrals for Topology $A$ in the region $0<t<1$}
\label{Sec6}

In this section we present our final result for the master integrals $f_8^{(2)}$, $f_9^{(2)}$ and $f_{10}$
in terms of iterated Eisenstein integrals. We focus in this section on the region $0<t<1$, and we explore the analytic continuation 
to other regions in the next section.

We start by discussing $f_8^{(2)}$. We can follow the steps outlined in the previous section and express all the eMPLs that appear in the analytic expression
for $f_8^{(2)}$ in eq.~\eqref{f8E3} in terms of iterated Eisenstein integrals. We find
\beq\label{eq.sol8All}
f_8^{(2)}(t) = \Psi_1(t)\,f_{8,U}^{(2)}(\tau(t))\,,
\eeq
with
\bea\label{eq.sol8}
f_{8,U}^{(2)}(\tau) & = & 16 \, \IEs{1, & \textbf{a}^{(3)}_{6,1,0}, & \textbf{a}^{(2)}_{6,1,0}}{\tau} + 16 \,\IEs{1, & \textbf{a}^{(3)}_{6,1,0}, & \textbf{a}^{(2)}_{6,1,3}}{\tau}  + 16\, \IEs{1, & \textbf{a}^{(3)}_{6,1,0}, & \textbf{a}^{(2)}_{6,2,0}}{\tau}  \nn \\
&& + 16\, \IEs{1, & \textbf{a}^{(3)}_{6,1,3}, & \textbf{a}^{(2)}_{6,1,0}}{\tau}  + 16\, \IEs{1, & \textbf{a}^{(3)}_{6,1,3}, & \textbf{a}^{(2)}_{6,1,3}}{\tau} + 16 \,\IEs{1, & \textbf{a}^{(3)}_{6,1,0}, & \textbf{a}^{(2)}_{6,2,0}}{\tau}  \nn \\
&& - 40\, \IEs{1, & \textbf{a}^{(3)}_{6,2,3}, & \textbf{a}^{(2)}_{6,1,0}}{\tau}  - 40\, \IEs{1, & \textbf{a}^{(3)}_{6,2,3}, & \textbf{a}^{(2)}_{6,1,3}}{\tau}  - 40 \,\IEs{1, & \textbf{a}^{(3)}_{6,2,3}, & \textbf{a}^{(2)}_{6,2,0}}{\tau}  \nn \\
&& + 4\,\log 3 \left(5 \, \IEs{1, & \textbf{a}^{(3)}_{6,2,3}}{\tau}   - 2\, \IEs{1, & \textbf{a}^{(3)}_{6,1,0}}{\tau} -2\, \IEs{1, & \textbf{a}^{(3)}_{6,1,3}}{\tau} \right)\nn \\
&& -i \operatorname{Cl}_2\left(\frac{\pi}{3}\right) \tau + \frac{6}{\pi} \operatorname{Im}G\left(0,1,1,e^{\frac{2 i \pi}{3}}\right) + \frac{2\pi^2}{27}\,,
\eea
where $\textrm{Cl}_2(x)$ denotes the Clausen function,
\beq
\textrm{Cl}_2(x) = \frac{i}{2}\left[\textrm{Li}_2(e^{-ix})-\textrm{Li}_2(e^{ix})\right]\,.
\eeq
The variable $\tau(t)$ is defined in eq.~\eqref{eq:tau_to_psi} and is we recalled here for convenience,
\be
\label{Tau}
\tau(t) = \frac{\Psi_2 (t)}{\Psi_1 (t)}.
\ee
Equation~\eqref{Tau} can be inverted, and we find~\cite{MaierME,Bloch:2013tra,Adams:2017ejb}
\be
\label{Haup}
t(\tau) = 9 \frac{\eta(\frac{\tau}{2})^4 \eta(3 \tau)^8}{\eta(\tau)^8 \eta(\frac{3\tau}{2})^4},
\ee
where $\eta(\tau)$ is the Dedekind $\eta$-function
\be
\eta(\tau) = q^{\frac{1 }{24}}\prod_{n = 1}^{\infty}(1 - q^n)\,, \qquad q=\exp(2\pi i\tau)\,.
\ee

Next, let us discuss $f_9^{(2)}$. Using its differential equation, we obtain
\be\label{eq:f9tof8}
f_9^{(2)}(t) = -\partial_tf_8^{(2)}(t) = -\Phi_1(t)\,f_{8,U}^{(2)}(\tau(t)) - \Psi_1(t)\mathcal{J}(t)\,\partial_{\tau}f_{8,U}^{(2)}(\tau(t))\,,
\ee
where $\Phi_1(t) = \partial_t\Psi_1(t)$ was defined in eq.~\eqref{eq:pMat} and $\mathcal{J}(t)$ denotes the Jacobian of the change of variables from $t$ to $\tau$,
\be
\label{Jac}
\mathcal{J}(t) = \partial_t\tau(t) = 
- \frac{48 i \pi}{(t- 9)(t - 1)t\Psi_1^2(t)}\,.
\ee
The derivative of $f_{8,U}^{(2)}$ with respect to $\tau$ can easily be carried out, as the iterated integrals in eq.~\eqref{eq.sol8} only depend on $\tau$ through the upper integration limit. We find
\bea\label{eq.sol9}
f_{9,U}^{(2)}(\tau)  & = & 2\pi i \,\partial_{\tau}f_{8,U}^{(2)}(\tau) \nn \\
&=&16\,  \IEs{\textbf{a}^{(3)}_{6,1,0}, & \textbf{a}^{(2)}_{6,1,0}}{\tau} + 16\, \IEs{\textbf{a}^{(3)}_{6,1,0}, & \textbf{a}^{(2)}_{6,1,3}}{\tau}  + 16 \,\IEs{\textbf{a}^{(3)}_{6,1,0}, & \textbf{a}^{(2)}_{6,2,0}}{\tau}  \nn \\
&& + 16\, \IEs{\textbf{a}^{(3)}_{6,1,3}, & \textbf{a}^{(2)}_{6,1,0}}{\tau}  + 16 \,\IEs{\textbf{a}^{(3)}_{6,1,3}, & \textbf{a}^{(2)}_{6,1,3}}{\tau} + 16 \,\IEs{\textbf{a}^{(3)}_{6,1,3}, & \textbf{a}^{(2)}_{6,2,0}}{\tau}  \nn \\
&& - 40\, \IEs{\textbf{a}^{(3)}_{6,2,3}, & \textbf{a}^{(2)}_{6,1,0}}{\tau}  - 40\, \IEs{\textbf{a}^{(3)}_{6,2,3}, & \textbf{a}^{(2)}_{6,1,3}}{\tau}  - 40 \,\IEs{\textbf{a}^{(3)}_{6,2,3}, & \textbf{a}^{(2)}_{6,2,0}}{\tau}  \nn \\
&& + 4\,\log 3 \left[5  \,\IEs{\textbf{a}^{(3)}_{6,2,3}}{\tau}   - 2 \,\IEs{\textbf{a}^{(3)}_{6,1,0}}{\tau} -2 \,\IEs{\textbf{a}^{(3)}_{6,1,3}}{\tau} \right]+2\pi\operatorname{Cl}_2\left(\frac{\pi}{3}\right).
\eea

Finally, we discuss the calculation of $f_{10}$. 
Aside from classical polylogarithms, the differential equation for $f_{10}$ 
in eq.~\eqref{EqM10} contains $f_8^{(2)}$ as an inhomogeneous term. In order to 
solve the differential equation we follow the strategy of ref.~\cite{Broedel:2019kmn}. We start by noting that we have expressed
$f_8^{(2)}$ in eq.~\eqref{eq.sol8All} in terms of iterated Eisenstein integrals for the congruence subgroup $\Gamma(6)$, for which
a spanning set is given by the functions in eq.~\eqref{eq:defAandB}. 
It turns out that there is a smaller set of modular forms that is sufficient to express the result for $f_8^{(2)}$, 
namely Eisenstein series for $\Gamma_1(6)$. 
In ref.~\cite{Broedel:2018rwm}
it was shown that a basis for $\mathcal{M}_n(\Gamma_1(6))$ is
\be
\label{G16b}
f_{n,p}(\tau) = t(\tau)^p \Psi_1(t(\tau))^n, \;\;\; 0 \leq p \leq n, \;\;\; f_{0,0}(\tau) = 1\,.
\ee
It is possible to write all polylogarithms that appear in eq.~\eqref{EqM10} in terms of iterated Eisenstein integrals for $\Gamma_1(6)$.
Here we only sketch the argument, and we refer to refs.~\cite{Broedel:2018rwm,Adams:2017ejb,Broedel:2019kmn} for details.
If we express the iterated integrals in eq.~\eqref{eq.sol8} in terms of the kernels in eq.~\eqref{G16b}, and we change variables from $\tau$ to $t$ using eq.~\eqref{Tau},
we can write eq.~\eqref{eq.sol8} in terms of iterated integrals in $t$, with integration kernels of the form
\beq
\frac{dt}{t-t_0}\,\Psi_1(t)^{n-2}\,,\qquad t_0\in \{0,1,9\}\,.
\eeq
This class of iterated integrals contains at the same time MPLs (for $n=2$) and iterated
Eisenstein integrals for $\Gamma_1(6)$ (through eq.~\eqref{G16b}). In other words, all contributions 
in eq.~\eqref{EqM10} can be expressed in terms of a unique class of iterated integrals,
the iterated Eisenstein integrals for $\Gamma_1(6)$,
and thus also in terms of Eisenstein series for $\Gamma(6)$. Let us consider as an example $G(0;t)$.
Using eq.~\eqref{Jac} it is possible to verify that the following identity holds
\be\label{eq:gtoiie}
\int_0^t \frac{dt}{t} = c + \frac{i}{48\pi}\left[\int_{i\infty}^{\tau}t^2(\tau)\Psi_1(t(\tau))^2d\tau - 10 \int_{i\infty}^{\tau} t(\tau)\Psi_1(t(\tau))^2d\tau +9 \int_{i\infty}^{\tau}\Psi_1(t(\tau))^2d\tau\right],
\ee
where $c$ is a boundary term associated with the regularisation of $G(0;t)$. From the $q$-expansion of the 
iterated integrals on the right-hand side of the above equation \cite{Broedel:2019kmn}, 
it is possible to write it in terms of iterated Eisenstein integrals :
\be
G(0;t) = \log 9 - 4\IEs{\textbf{a}^{(2)}_{6,1,0}}{\tau} - 4\IEs{\textbf{a}^{(2)}_{6,1,3}}{\tau} - 4\IEs{\textbf{a}^{(2)}_{6,2,0}}{\tau},
\ee 
with $c = \log 9$.

After this step, the inhomogeneous term of eq.~\eqref{EqM10} only involves Eisenstein series and iterated Eisenstein integrals,
and the differential equation for $f_{10}$ can easily be solved. We find
\bea\label{eq.sol10}
f_{10}(t) & = & 384 \, \IEs{\textbf{a}^{(3)}_{6,1,0}, & 1, & \textbf{a}^{(3)}_{6,1,0}, & \textbf{a}^{(2)}_{6,1,0}}{\tau} + 384 \, \IEs{\textbf{a}^{(3)}_{6,1,0}, & 1, & \textbf{a}^{(3)}_{6,1,0}, & \textbf{a}^{(2)}_{6,1,3}}{\tau} \nn \\
&& + 384\,  \IEs{\textbf{a}^{(3)}_{6,1,0}, & 1, & \textbf{a}^{(3)}_{6,1,0}, & \textbf{a}^{(2)}_{6,2,0}}{\tau}  + 384 \, \IEs{\textbf{a}^{(3)}_{6,1,0}, & 1, & \textbf{a}^{(3)}_{6,1,3}, & \textbf{a}^{(2)}_{6,1,0}}{\tau} \nn \\
&& + 384\,\IEs{\textbf{a}^{(3)}_{6,1,0}, & 1, & \textbf{a}^{(3)}_{6,1,3}, & \textbf{a}^{(2)}_{6,1,3}}{\tau} + 384\, \IEs{\textbf{a}^{(3)}_{6,1,0}, & 1, & \textbf{a}^{(3)}_{6,1,3}, & \textbf{a}^{(2)}_{6,2,0}}{\tau}\nn \\
&& -960\, \IEs{\textbf{a}^{(3)}_{6,1,0}, & 1, & \textbf{a}^{(3)}_{6,2,3}, & \textbf{a}^{(2)}_{6,1,0}}{\tau} -960\, \IEs{\textbf{a}^{(3)}_{6,1,0}, & 1, & \textbf{a}^{(3)}_{6,2,3}, & \textbf{a}^{(2)}_{6,1,3}}{\tau}\nn \\
&& -960\,\IEs{\textbf{a}^{(3)}_{6,1,0}, & 1, & \textbf{a}^{(3)}_{6,2,3}, & \textbf{a}^{(2)}_{6,2,0}}{\tau} + 384\,\IEs{\textbf{a}^{(3)}_{6,1,3}, & 1, & \textbf{a}^{(3)}_{6,1,0}, & \textbf{a}^{(2)}_{6,1,0}}{\tau} \nn \\
&& + 384\,\IEs{\textbf{a}^{(3)}_{6,1,3}, & 1, & \textbf{a}^{(3)}_{6,1,0}, & \textbf{a}^{(2)}_{6,1,3}}{\tau}  + 384\,\IEs{\textbf{a}^{(3)}_{6,1,3}, & 1, & \textbf{a}^{(3)}_{6,1,0}, & \textbf{a}^{(2)}_{6,2,0}}{\tau}  \nn \\
&& + 384\,\IEs{\textbf{a}^{(3)}_{6,1,3}, & 1, & \textbf{a}^{(3)}_{6,1,3}, & \textbf{a}^{(2)}_{6,1,0}}{\tau} + 384\,\IEs{\textbf{a}^{(3)}_{6,1,3}, & 1, & \textbf{a}^{(3)}_{6,1,3}, & \textbf{a}^{(2)}_{6,1,3}}{\tau} \nn \\ 
&& + 384\,\IEs{\textbf{a}^{(3)}_{6,1,3}, & 1, & \textbf{a}^{(3)}_{6,1,3}, & \textbf{a}^{(2)}_{6,2,0}}{\tau}  - 960\,\IEs{\textbf{a}^{(3)}_{6,1,3}, & 1, & \textbf{a}^{(3)}_{6,2,3}, & \textbf{a}^{(2)}_{6,1,0}}{\tau}  \nn \\
&& - 960\,\IEs{\textbf{a}^{(3)}_{6,1,3}, & 1, & \textbf{a}^{(3)}_{6,2,3}, & \textbf{a}^{(2)}_{6,1,3}}{\tau} - 960\,\IEs{\textbf{a}^{(3)}_{6,1,3}, & 1, & \textbf{a}^{(3)}_{6,2,3}, & \textbf{a}^{(2)}_{6,2,0}}{\tau} \nn \\
&& +192\,\IEs{\textbf{a}^{(3)}_{6,2,3}, & 1, & \textbf{a}^{(3)}_{6,1,0}, & \textbf{a}^{(2)}_{6,1,0}}{\tau} +192\,\IEs{\textbf{a}^{(3)}_{6,2,3}, & 1, & \textbf{a}^{(3)}_{6,1,0}, & \textbf{a}^{(2)}_{6,1,3}}{\tau} \nn \\
&& +192\,\IEs{\textbf{a}^{(3)}_{6,2,3}, & 1, & \textbf{a}^{(3)}_{6,1,0}, & \textbf{a}^{(2)}_{6,2,0}}{\tau}  +192\,\IEs{\textbf{a}^{(3)}_{6,2,3}, & 1, & \textbf{a}^{(3)}_{6,1,3}, & \textbf{a}^{(2)}_{6,1,0}}{\tau}  \nn \\
&& +192\,\IEs{\textbf{a}^{(3)}_{6,2,3}, & 1, & \textbf{a}^{(3)}_{6,1,3}, & \textbf{a}^{(2)}_{6,1,3}}{\tau}   +192\IEs{\textbf{a}^{(3)}_{6,2,3}, & 1, & \textbf{a}^{(3)}_{6,1,3}, & \textbf{a}^{(2)}_{6,2,0}}{\tau}   \nn \\
&& - 480\,\IEs{\textbf{a}^{(3)}_{6,2,3}, & 1, & \textbf{a}^{(3)}_{6,2,3}, & \textbf{a}^{(2)}_{6,1,0}}{\tau}   - 480\,\IEs{\textbf{a}^{(3)}_{6,2,3}, & 1, & \textbf{a}^{(3)}_{6,2,3}, & \textbf{a}^{(2)}_{6,1,3}}{\tau}  \nn \\
&& - 480\,\IEs{\textbf{a}^{(3)}_{6,2,3}, & 1, & \textbf{a}^{(3)}_{6,2,3}, & \textbf{a}^{(2)}_{6,2,0}}{\tau}  + \log 3 \left[480\,\IEs{\textbf{a}^{(3)}_{6,1,0}, & 1, & \textbf{a}^{(3)}_{6,2,3}}{\tau} \right.  \nn \\
&& \left. - 192\, \IEs{\textbf{a}^{(3)}_{6,1,0}, & 1, & \textbf{a}^{(3)}_{6,1,3}}{\tau} -192\, \IEs{\textbf{a}^{(3)}_{6,1,0}, & 1, & \textbf{a}^{(3)}_{6,1,0}}{\tau}  \right. \nn \\
&& \left. -192\,\IEs{\textbf{a}^{(3)}_{6,1,3}, & 1, & \textbf{a}^{(3)}_{6,1,0}}{\tau} - 192 \, \IEs{\textbf{a}^{(3)}_{6,1,3}, & 1, & \textbf{a}^{(3)}_{6,1,3}}{\tau} \right. \nn \\ 
&& \left. + 480\,\IEs{\textbf{a}^{(3)}_{6,1,3}, & 1, & \textbf{a}^{(3)}_{6,2,3}}{\tau} - 96 \,\IEs{\textbf{a}^{(3)}_{6,2,3}, & 1, & \textbf{a}^{(3)}_{6,1,0}}{\tau} \right.  \nn \\
&& \left. - 96 \,\IEs{\textbf{a}^{(3)}_{6,2,3}, & 1, & \textbf{a}^{(3)}_{6,1,3}}{\tau} + 240 \, \IEs{\textbf{a}^{(3)}_{6,2,3}, & 1, & \textbf{a}^{(3)}_{6,2,3}}{\tau}\right] \nn \\
&& + \left[\frac{72}{\pi}\operatorname{Im} G\!\left(0,1,1,e^{\frac{2 i \pi}{3}}\right) +\frac{8 \pi^2}{9} \right]\left[2\, \IEs{\textbf{a}^{(3)}_{6,1,0}}{\tau} + 2 \, \IEs{\textbf{a}^{(3)}_{6,1,3}}{\tau} +  \IEs{\textbf{a}^{(3)}_{6,2,3}}{\tau} \right]  \nn \\
&& +24\,\pi\,\operatorname{Cl}_2\left(\frac{\pi}{3}\right)\left[2\, \IEs{\textbf{a}^{(3)}_{6,1,0},&1}{\tau} + 2 \, \IEs{\textbf{a}^{(3)}_{6,1,3},&1}{\tau} +  \IEs{\textbf{a}^{(3)}_{6,2,3}&,1}{\tau} \right] \nn \\
&& + 2G\left(0,1,1,0;t\right) -G\left(1,0,1,0;t\right) + \frac{\pi^2}{6} G\left(1,0;t\right) - \frac{\pi^2}{3} G\left(0,1;t\right) - \frac{22\zeta_3}{3} G(1;t) \nn \\
&& - 6 \operatorname{Cl}_2\left(\frac{\pi}{3}\right)^2 -\frac{\pi^4}{24}. 
\eea

Equations~\eqref{eq.sol8},~\eqref{eq.sol9} and~\eqref{eq.sol10} are among the main results of this paper. 
They express all elliptic master integrals of Topology $A$ in terms of a class of special functions that is well studied in both the mathematics and physics literature, namely eMPLs and iterated Eisenstein integrals for the congruence subgroup $\Gamma(6)$. The discussion from the previous paragraph shows that we can even restrict the analysis to the larger congruence subgroup $\Gamma_1(6)$, at least through finite terms in the Laurent expansion in $\eps$. This shows that all master integrals for Topology $A$ can be expressed in terms of exactly the same class of functions as the well-known sunrise, kite and banana integrals with three or four equal masses. This extends the observation of Section~\ref{sec:notAndConv} that $f_8^{(2)}$ and $f_9^{(2)}$ satisfy the same homogeneous differential equation as the sunrise graph. In particular, we find that there is no need to introduce new classes of transcendental functions beyond those already encountered for the sunrise and kite graphs. This is at variance with the analytic results for Topology $A$ of refs.~\cite{Ablinger:2017bjx,Blumlein:2018aeq}, where new classes of functions were introduced.

Since the integrals considered here and the sunrise graph seem to be so closely related, let us comment on the elliptic curves that appear in the computation of the two integrals. As already mentioned in Section~\ref{sec:notAndConv}, the second order differential operator in eq.~\eqref{Dt2} that describes the homogeneous solution is identical for the two sets of integrals. The solutions of eq.~\eqref{Dt2} are the two periods $\Psi_1(t)$ and $\Psi_2(t)$ of the family of elliptic curves parametrised by $t$. While in the case of the integrals considered here this curve is most naturaly defined by a cubic polynomial (cf.~eq.~\eqref{eq:curveRho}), the curve obtained from the Feynman parameter integral for the sunrise is defined by a quartic polynomial, cf.~e.g.~refs.~\cite{Bloch:2013tra,MullerStach:2011ru,Broedel:2017siw}. There is no contradiction: the same elliptic curve may be represented as the zero set of different polynomial equations. An invariant that uniquely distinguishes different elliptic curves is its $j$-invariant. The $j$-invariant of the family of elliptic curves in eq.~\eqref{eq:curveRho} is
\beq\label{eq:jInv}
j(t) =\frac{(t-3)^3 (t ((t-9) t+3)-3)^3}{1728 (t-9)(t-1)^3 t^2}\,.
\eeq
It is easy to check that eq.~\eqref{eq:jInv} agrees with the $j$-invariant for the family of elliptic curves obtained from the Feynman parametrisation of the equal-mass sunrise graph, with {$t=\frac{p^2}{m^2}$~\cite{Bloch:2013tra}}. This shows that indeed the elliptic curves obtained from the Feynman parameter integrals of the sunrise integrals and the integrals considered here are identical. 

Finally, let us make a comment about the analytic structure of our results. We see that we can cast our results in the form,
 \be
 \label{SemiDec}
\left(\begin{array}{c} f_8^{(2)}(t) \\ f_9^{(2)}(t)   \\ f_{10}(t) \end{array}\right)
= \mathcal{S}(t) \left(\begin{array}{c} f_{8,U}^{(2)}(t) \\ f_{9,U}^{(2)}(t)   \\ f_{10,U}(t) \end{array}\right),
 \ee
 where $f_{8,U}^{(2)}(t)$, $f_{9,U}^{(2)}(t)$ and $f_{10,U}(t)=f_{10}(t)$ are defined in
 eqs.~\eqref{eq.sol8},~\eqref{eq.sol9} and~\eqref{eq.sol10}. The
 matrix $\mathcal{S}$ is given by
\be
\mathcal{S}(t) = \left(\begin{array}{ccc}
\Psi_1(t) & 0 & 0 \\
-\Phi_1(t) & \frac{24}{(t - 9)(t- 1)t\Psi_1(t)} & 0 \\
0 & 0 & 1
\end{array}\right).
\ee
This form matches precisely the structure of elliptic Feynman integrals conjectured in ref.~\cite{Broedel:2018qkq}.
In particular, we see that the functions $f_{8,U}^{(2)}(t)$, $f_{9,U}^{(2)}(t)$ and $f_{10,U}(t)$ are pure functions in the sense of ref.~\cite{Broedel:2018qkq},
 and they have uniform transcendental weight two, three and four respectively.


\section{Analytic continuation and numerical evaluation}
\label{Sec7}

\subsection{Analytic continuation}

As mentioned in Section \ref{sec:notAndConv}, for the calculation of the $\rho$ parameter we do not only
need the integrals from Topology $A$ (see fig.~\ref{fig:allTopo}),
but also those from Topology $B$, obtained by exchanging 
$m_1$ and $m_2$. These integrals can equivalently be obtained
by analytically continuing Topology $A$ to the region $t>1$. 
The analytic continuation of the non-elliptic
integrals can be done using standard techniques.
In this section we discuss the analytic continuation of the elliptic integrals $f_8^{(2)}$, $f_9^{(2)}$ and $f_{10}$.

The analytic continuation will be done following the steps described in ref.~\cite{Duhr:2019rrs}. We start by discussing the 
analytic continuation of the homogeneous solutions in eq.~\eqref{periods}.
Given the singularities in the differential equation in eq.~\eqref{Dt2}, 
there are four kinematic regions to consider: $t<0$, $0<t<1$, $1<t<0$, $t>9$. 
The homogeneous solution in eq.~\eqref{periods} is well behaved in the second region, by which we mean
that they are local solutions to the differential equation that are respectively real and imaginary for $t\in[0,1]$. 
The first step in
the analytic continuation procedure is to obtain similarly well behaved solutions in the other
three regions. Since the differential equation in eq.~\eqref{Dt2} is the same as that of the 
sunrise integral, we can simply reuse the results of refs.~\cite{Remiddi:2016gno,Bogner:2017vim}:
\begin{itemize}
\item{Region 1, $0 < t < 1$:
\be
\label{R1}
\tilde{\Psi}_1^{(0,1)}(t) = \frac{8\,\K(\lambda(t))}{\sqrt{(3- \sqrt{t})(1 + \sqrt{t})^3}}, \;\; 
\tilde{\Psi}_2^{(0,1)}(t) = \frac{16 i \,\K(1 - \lambda(t))}{\sqrt{(3- \sqrt{t})(1 + \sqrt{t})^3}}\,;
\ee}
\item{Region 2, $1 < t < 9$:
\be
\label{R2}
\tilde{\Psi}_1^{(1,9)}(t) = \frac{2}{t^{\frac{1}{4}}}\K\left(\frac{1}{\lambda(t)}\right), \;\; 
\tilde{\Psi}_2^{(1,9)}(t) = \frac{4 i }{t^{\frac{1}{4}}}\K\left(1 - \frac{1}{\lambda(t)}\right); 
\ee}
\item{Region 3, $t > 9$:
\be
\label{R3}
\tilde{\Psi}_1^{(9,\infty)}(t) = \frac{8\,\K\left(\lambda_9(t)\right)}{\sqrt{(3 + \sqrt{t})(\sqrt{t} - 1)^3}}, \;\; 
\tilde{\Psi}_2^{(9,\infty)}(t) = \frac{16 i\,\K\left(1 - \lambda_9(t)\right)}{\sqrt{(3 + \sqrt{t})(\sqrt{t} - 1)^3}}\,; 
\ee}
\item{Region 4, $t < 0$:
\be
\label{R4}
\tilde{\Psi}_1^{(-\infty,0)}(t) = \frac{8\,\K\left(\lambda_0(t)\right)}{\left((t - 9)(t - 1)^3\right)^{\frac{1}{4}}}, \;\;
\tilde{\Psi}_2^{(-\infty,0)}(t) = \frac{4 i\,\K\left(1-\lambda_0(t)\right)}{\left((t-9)(t - 1)^3\right)^{\frac{1}{4}}}\,; 
\ee}
\end{itemize}
where $\lambda(t)$ was defined in eq.~\eqref{eq:lambda_def} while $\lambda_9(t)$ and $\lambda_0(t)$ are given by:
\begin{align}\begin{split}
\lambda_9(t) =& \frac{(\sqrt{t} - 3)(\sqrt{t} + 1)^3}{(3 + \sqrt{t})(\sqrt{t} - 1)^3}\,, \;\;\; \\
\lambda_0(t) =& \frac{(t - 9)\left(3 - \sqrt{(t - 9)(t - 1)} + t(6 - t + \sqrt{(t - 9)(t - 1)})\right)}{2\left((t - 9)(t - 1)\right)^{\frac{3}{2}}}\,.
\end{split}\end{align}

The functions $\tilde{\Psi}^{(a,b)}_j$ are local solutions to the differential equation in eq.~\eqref{Dt2}, but they do not extend individually to global solutions that define
analytic functions with at most logarithmic singularities at the regular singular points of eq.~\eqref{Dt2}. We can, however, construct a set of solutions
with the desired analytic properties by patching together the local solutions in the right way. This leads to the correct analytic continuation of the functions in eq.~\eqref{periods} to all values of $t$, given by the piecewise definition
\begin{equation}\label{eq:fullSolHomo}
\left(\Psi_1(t), \Psi_2(t)\right) = \left\{
\begin{array}{ll}
	\left(\tilde{\Psi}_1^{(0,1)}(t), \tilde{\Psi}_2^{(0,1)}(t)\right)\,
	\left(\begin{array}{cc}1 & 0 \\ 0 & 1\end{array}\right)&\qquad 
	\text{for }0\leq t<1\,,\\[4mm]
	\left(\tilde{\Psi}_1^{(1,9)}(t), \tilde{\Psi}_2^{(1,9)}(t)\right)\,
	\left(\begin{array}{cc}1 & 0 \\ \frac{3}{2} & 1\end{array}\right)&\qquad 
	\text{for }1\leq t<9\,,\\[4mm]
	\left(\tilde{\Psi}_1^{(9,\infty)}(t), \tilde{\Psi}_2^{(9,\infty)}(t)\right)\,
	\left(\begin{array}{cc}-2 & -2 \\ \frac{3}{2} & 1\end{array}\right)&\qquad 
	\text{for }9\leq t<\infty\,,\\[4mm]
	\left(\tilde{\Psi}_1^{(-\infty,0)}(t), \tilde{\Psi}_2^{(-\infty,0)}(t)\right)\,
	\left(\begin{array}{cc}1 & 1 \\ 0 & 2\end{array}\right)&\qquad 
	\text{for } -\infty<t<0\,.
\end{array}\right.
\end{equation}
Having obtained the correct analytic continuation of the homogeneous solutions to all values of $t$, we can also extend the definition of $\tau(t)$ in eq.~\eqref{eq:tau_to_psi} to arbitrary $t$, by replacing $\Psi_j(t)$ in eq.~\eqref{eq:tau_to_psi} by their piecewise definition in eq.~\eqref{eq:fullSolHomo}. For the convenience of the reader, we reproduce here the definition of $\tau(t)$ as a function of the homogeneous solutions:
\be\label{eq:tauAllRegions}
\tau(t) = \frac{\Psi_2(t)}{\Psi_1(t)}\,.
\ee

With these definitions we can immediately evaluate the integrals $f_8^{(2)}$, $f_9^{(2)}$ and $f_{10}$ for all real values of $t$, for both Topology $A$ and $B$.
Indeed, we can insert the global definitions of $\Psi_j$ and $\tau$ from eqs.~\eqref{eq:fullSolHomo} and~\eqref{eq:tauAllRegions} into eq.~\eqref{SemiDec}. 
Each iterated Eisenstein integral admits a
$q$-expansion (see eq.~\eqref{Qexp}), which is guaranteed to converge because $\textrm{Im }\tau(t)>0$ for all values of $t$. 
Therefore, we can obtain numerical results for all the integrals in eq.~\eqref{SemiDec} for arbitrary $t$.
The convergence of the $q$-expansion, however, might
be very slow: it is controlled by the size of the imaginary part of $\tau$
which, depending on the value of $t$, might be very small. In the
next section we address this shortcoming.

\subsection{Numerical evaluation of iterated integrals of modular forms}

In this section we discuss how we can speed up the numerical convergence of 
the $q$ expansion of the iterated Eisenstein integrals.
Indeed, for physical applications it is desirable to have representations
for the master integrals that can be evaluated efficiently. This may however 
not be the case for our results, which are simply obtained by inserting the analytic continuation of the 
homogenous solutions into eqs.~\eqref{eq:tauAllRegions} and \eqref{SemiDec},
without care for the convergence properties of the expression we obtain.

To make our point more concrete, let us consider the case where $t$ takes its SM value $t_{\textrm{phys}}$ (cf.~eq.~\eqref{eq:tPhys}). 
For Topology $A$, we then find $\tau(t_{\textrm{phys}}) \simeq i\,3.07$. 
This corresponds to a value of $q_6 = \exp(\pi i\tau(t)/3) \simeq0.04$ in the $q$-expansion of the 
iterated Eisenstein integrals (see eq.~\eqref{Qexp}) and so all integrals admit a fast converging $q$-expansion. 
For Topology $B$, however, the integrals are evaluated at $\tau(1/t_{\textrm{phys}}) \simeq 0.94+i\,0.13$, which gives $|q_6| \simeq 0.88$.
The iterated Eisenstein integrals computed with $\tau(t_\text{phys})$ will then converge 
one order of magnitude faster than those computed with $\tau(1/t_\text{phys})$.

The convergence of the $q$-expansion can be substantially accelerated, 
for instance following the procedure of ref.~\cite{Duhr:2019rrs} which we now briefly summarise. 
In a nutshell, the idea is to find a transformation 
$\gamma\in\text{SL}(2,\mathbb{Z})$ that maximises the imaginary part of $\tau$
and thus the speed of the convergence of the $q$-expansion. For example, we can map $\tau$ to the so-called \emph{fundamental domain} of 
$\text{SL}(2,\mathbb{Z})$, defined as
\be
\mathcal{F} = \left\{\tau \in \mathbb{H}\, : \, -\frac{1}{2}\leq \operatorname{Re}(\tau) \, < \, \frac{1}{2} \, \operatorname{and} \, \vert \tau \vert \, > \, 1 \right\} \bigcup \left\{\tau \in \mathbb{H} \, : \, \operatorname{Re}(\tau) \, \leq \, 0 \, \operatorname{and} \, \vert \tau \vert \, = \, 1\right\}.
\ee
All $\tau\in\mathcal{F}$ have
$\operatorname{Im}(\tau)\geq \sqrt{3}/2$ and for every value of $t$, i.e., for every $\tau(t) \in \mathbb{H}$, there is a $\gamma_t\in\text{SL}(2,\mathbb{Z})$ such that $\gamma_t^{-1}\cdot\tau(t)\in\mathcal{F}$.

While we might naively expect that the map to the fundamental domain in each of the four different regions for $t$ 
in eq.~\eqref{eq:fullSolHomo} is given by a single transformation, or in other words that the function $\gamma_t$ is constant on each of the four regions in eq.~\eqref{eq:fullSolHomo}, this is in fact not the case. 
In ref.~\cite{Duhr:2019rrs} it was observed that the points where $\gamma_t$ changes are a subset
of the (real) solutions to the equations 
\begin{equation}
	j(t)=0\,,\qquad
	j(t)=1\,,\qquad\text{or}\qquad
	j(t)=\pm\infty\,,
\end{equation}
where $j(t)$ is the  $j$-invariant introduced in eq.~\eqref{eq:jInv}.
Note that the solutions to $j(t)=\pm\infty$ are the singularities of
the differential equation \eqref{eq:diffEll}, which define the different
regions for analytic continuation, see eq.~\eqref{eq:fullSolHomo}.
By inspection, we then find that the points where $\gamma_t$ changes are
\begin{align}\begin{split}\label{eq:tiReg}
t_1 &= 3-2\sqrt{3}\,,\quad  t_2 = 0\,,\quad  t_3 =  3+\sqrt{3}-\sqrt{9+6 \sqrt{3}}\,,\quad t_4 = 1\,, \\
t_5 &= 3 \,,\quad t_6 = 3+2 \sqrt[3]{2}+2^{5/3}\,,\quad t_7 = 3+\sqrt{3}+\sqrt{9+6 \sqrt{3}}\,,\quad t_8=\infty\,.
\end{split}\end{align}

For each of the eight regions bounded by the values of $t$
in eq.~\eqref{eq:tiReg} we can identify a modular transformation $\gamma_t$ 
whose inverse maps $\tau(t)$ to the fundamental domain.
This can be done algorithmically~\cite{Duhr:2019rrs}. Through this
procedure we obtain solutions for the master integrals $f_8^{(2)}(t)$, 
$f_9^{(2)}(t)$ and $f_{10}(t)$ that can be efficiently evaluated for
all real values of $t$. This approach also allows one to obtain fast converging expansions for 
the master integrals in Topology $B$ starting from the expressions for Topology $A$,
since one is obtained from the other by the replacement $t\to 1/t$.

Before we delve into the details of how we obtain an expression for the masters
$f_8^{(2)}(t)$, $f_9^{(2)}(t)$ and $f_{10}(t)$ that is tailored for efficient numerical
evaluation, we start with a comment about the notation we will use. We will
proceed by analogy with what we did in the previous section for the analytic
continuation, where the global solutions have piecewise definitions,
constructed from local solutions that are well defined in each region (see e.g.~eqs.~\eqref{eq:fullSolHomo}~and~\eqref{eq:tauAllRegions}). Here we will
denote by an index $A$ the quantities that are associated with Topology $A$
and admit a piecewise definition
constructed from local representations, which can be efficiently evaluated
in each of the eight regions defined by the points in eq.~\eqref{eq:tiReg}.
For instance, in analogy with eq.~\eqref{SemiDec}, we define
 \be\label{eq:allRegA}
\vec{f}_A(t)=\left(\begin{array}{c} f_{8}^{(2),A}(t) \\ f_{9}^{(2),A}(t)   \\ f^{A}_{10}(t) \end{array}\right)
= \mathcal{S}_A(t) \left(\begin{array}{c} f_{8,U}^{(2),A}(t) \\ f_{9,U}^{(2),A}(t)   \\ f^{A}_{10,U}(t) \end{array}\right),
 \ee
where all quantities have a piecewise definition, and the pure integrals
$f_{8,U}^{(2),A}(t)$, $f_{9,U}^{(2),A}(t)$  and $f^{A}_{10,U}(t)$ are evaluated
at $\tau_A(t)$, which itself admits a piecewise definition.
The remainder of this section is devoted to giving explicit representations
for these quantities, at least for the regions that are relevant for physical applications.

We first construct $\tau_A(t)\in\mathcal{F}$, defined as the image of 
the $\tau(t)$ given in eq.~\eqref{eq:tauAllRegions} under the map 
to the fundamental domain,
\begin{equation}
	\tau_A(t)=\gamma_t^{-1}\cdot\tau(t)\,.
\end{equation}
In accordance with the previous discussion, $\tau_A(t)$ admits a piecewise definition:
to each of the eight regions bounded by the points $t_j$ in eq.~\eqref{eq:tiReg}
corresponds a different matrix $\gamma_t$. We write
\begin{align}\label{eq:gammaTAll}
	\gamma_t=\gamma^{(j,j+1)} \text{~~~for~~~} t\in[t_j,t_{j+1}]\,,
\end{align}
where the indices are understood to be cyclically defined and
\begin{align}\begin{split}\label{eq:gammaTAllMat}
	\gamma^{(1,2)}=\left(
	\begin{array}{cc}
	 1 & 1 \\
	 0 & 1 \\
	\end{array}
	\right)\,,\quad
	\gamma^{(2,3)}=\left(
	\begin{array}{cc}
	 1 & 0 \\
	 0 & 1 \\
	\end{array}
	\right)\,,\quad
	\gamma^{(3,4)}=\left(
	\begin{array}{cc}
	 0 & -1 \\
	 1 & 0 \\
	\end{array}
	\right)\,,\quad
	\gamma^{(4,5)}=\left(
	\begin{array}{cc}
	 0 & -1 \\
	 1 & -2 \\
	\end{array}
	\right)\,,\\
	\gamma^{(5,6)}=\left(
	\begin{array}{cc}
	 1 & 1 \\
	 1 & 2 \\
	\end{array}
	\right)\,,\quad
	\gamma^{(6,7)}=\left(
	\begin{array}{cc}
	 2 & -1 \\
	 3 & -1 \\
	\end{array}
	\right)\,,\quad
	\gamma^{(7,8)}=\left(
	\begin{array}{cc}
	 1 & 2 \\
	 1 & 3 \\
	\end{array}
	\right)\,,\quad
	\gamma^{(8,1)}=
	\left(
	\begin{array}{cc}
	 1 & -1 \\
	 1 & 0 \\
	\end{array}
	\right)\,.
\end{split}\end{align}

While we gave an explicit definition for $\tau_A(t)$ for all values of $t$, doing the same
for the different quantities in eq.~\eqref{eq:allRegA} would lead us towards a very lengthy
and repetitive discussion. Instead, we give these expression as ancillary files and
focus our discussion in the remainder of this section on the two regions that are relevant for the
calculation of the $\rho$ parameter: $t\in[t_2,t_3]$ and $t\in[t_7,t_8]$.

Note that $\gamma^{(2,3)}$ is the identity matrix.
The value of $\tau(t)$ of eq.~\eqref{eq:tauAllRegions} is thus already in the fundamental domain for $t\in[t_2,t_3]$, and the integrals
defined in eq.~\eqref{SemiDec} are already in a representation with the best possible convergence properties.
In particular, we have
\be\begin{split}
	\left(
	\begin{array}{c}
	 \Psi_2^A(t) \\
	 \Psi_1^A(t)\\
	\end{array}
	\right)
	=\left(
	\begin{array}{c}
	 \Psi_2(t) \\
	 \Psi_1(t)\\
	\end{array}
	\right)\textrm{~~and~~}
	\vec{f}_A(t)=\left(\begin{array}{c} f_8^{(2)}(t) \\ f_9^{(2)}(t)   \\ f_{10}(t) \end{array}\right)
	\textrm{~~for~~} t\in[t_2,t_3]\,,
\end{split}\ee
with $\Psi_j(t)$ as defined in eq.~\eqref{eq:fullSolHomo}, and $f_8^{(2)}(t)$, $f_9^{(2)}(t)$ and $f_{10}(t)$
as in eq.~\eqref{SemiDec}.

Let us now discuss the region $t\in[t_7,t_8]$. It follows from eq.~\eqref{eq:gammaTAll} that
\begin{equation}\label{eq:tauALT}
	\tau_A(t) =\frac{3\tau(t)-2}{1-\tau(t)} \textrm{~~~for~~~} t\in[t_7,t_8] \,.
\end{equation}
Consistently, we have
\be\begin{split}
\label{PeriodsLT}
	\left(
	\begin{array}{c}
	 \Psi_2^A(t) \\
	 \Psi_1^A(t)\\
	\end{array}
	\right)
	=-i\left(
	\begin{array}{cc}
	 3 & -2 \\
	 -1 & 1 \\
	\end{array}
	\right)
	\left(
	\begin{array}{c}
	 \Psi_2(t) \\
	 \Psi_1(t)\\
	\end{array}
	\right)
	\textrm{~~~for~~~} t\in[t_7,t_8]\,,
\end{split}\ee
with the $\Psi_j(t)$ as defined in eq.~\eqref{eq:fullSolHomo}.
The factor of $-i$ is purely conventional and introduced so that $\Psi_1^A(t)$ is real 
and $\Psi_2^A(t)$ is imaginary also for $t\in[t_7,t_8]$.
We then find $\tau_A(1/t_\text{phys})\simeq i\,6.59$, which corresponds to $q_6 \simeq 0.001$.
Hence, if for $t\in[t_7,t_8]$ we express all iterated integrals in eq.~\eqref{SemiDec} in terms of iterated Eisenstein integrals 
evaluated at $\tau_A(t)$, as given in eq.~\eqref{eq:tauALT}, then we obtain fast converging $q$-expansions for the elliptic master integrals.
The corresponding transformations can be worked out using the algorithms presented in ref.~\cite{Duhr:2019rrs}. 
We then find that, in this region, the matrix $\mathcal{S}^A(t)$ of eq.~\eqref{eq:allRegA} is given by
\begin{equation}
\mathcal{S}_A(t) = \left(
\begin{array}{ccc}
 \Psi_1^A(t) & 0 & 0 \\
 -\Phi_1^A(t) & -\frac{24 }{(t-9) (t-1) t
   \Psi_1^A(t)} & 0 \\
 0 & 0 & 1 \\
\end{array}
\right)\textrm{~~~for~~~} t\in[t_7,t_8]\,.
\end{equation}
For $t\in[t_7,t_8]$, the pure integrals $f_{8,U}^{(2),A}(t)$, $f_{9,U}^{(2),A}(t)$  and $f^{A}_{10,U}(t)$ are:
\begin{align}\begin{split}\label{eq:f8ua}
f_{8,U}^{(2),A}(t) =\, & 40\, \IEs{1, & \textbf{b}^{(3)}_{6,0,1}, &  \textbf{a}^{(2)}_{6,0,1}}{\tau_A(t)} 
- 40\, \IEs{1, & \textbf{b}^{(3)}_{6,0,1}, &  \textbf{a}^{(2)}_{6,0,2}}{\tau_A(t)} \\
&+ 16 \IEs{1, & \textbf{b}^{(3)}_{6,3,1}, &  \textbf{a}^{(2)}_{6,0,1}}{\tau_A(t)}  
- 16\, \IEs{1, & \textbf{b}^{(3)}_{6,3,1}, &  \textbf{a}^{(2)}_{6,0,2}}{\tau_A(t)} \\
&- 16\, \IEs{1, & \textbf{b}^{(3)}_{6,3,2}, &  \textbf{a}^{(2)}_{6,0,1}}{\tau_A(t)} 
+ 16\, \IEs{1, & \textbf{b}^{(3)}_{6,3,2}, &  \textbf{a}^{(2)}_{6,0,2}}{\tau_A(t)}  \\
& - \frac{i \pi^2}{12}\tau_A(t) + \frac{\zeta_3}{\pi},
\end{split}\end{align}
\begin{align}\begin{split}
f_{9,U}^{(2),A}(t) = & 40 \,\IEs{\textbf{b}^{(3)}_{6,0,1}, &  \textbf{a}^{(2)}_{6,0,1}}{\tau_A(t)} 
- 40\,\IEs{\textbf{b}^{(3)}_{6,0,1}, &  \textbf{a}^{(2)}_{6,0,2}}{\tau_A(t)} \\
&+ 16\,\IEs{\textbf{b}^{(3)}_{6,3,1}, &  \textbf{a}^{(2)}_{6,0,1}}{\tau_A(t)}  
- 16\,\IEs{\textbf{b}^{(3)}_{6,3,1}, &  \textbf{a}^{(2)}_{6,0,2}}{\tau_A(t)} \\
&- 16\,\IEs{\textbf{b}^{(3)}_{6,3,2}, &  \textbf{a}^{(2)}_{6,0,1}}{\tau_A(t)} 
+ 16\,\IEs{\textbf{b}^{(3)}_{6,3,2}, &  \textbf{a}^{(2)}_{6,0,2}}{\tau_A(t)}+\frac{ \pi^3}{6},
\end{split}\end{align}
\begin{align}
f_{10,U}^{A}(t) =\, & 480\,\IEs{\textbf{b}^{(3)}_{6,0,1}, & 1, & \textbf{b}^{(3)}_{6,0,1}, &  \textbf{a}^{(2)}_{6,0,2}}{\tau_A(t)} 
- 480\,\IEs{\textbf{b}^{(3)}_{6,0,1}, & 1, & \textbf{b}^{(3)}_{6,0,1}, &  \textbf{a}^{(2)}_{6,0,1}}{\tau_A(t)}\nn  \\
& - 192\,\IEs{\textbf{b}^{(3)}_{6,0,1}, & 1, & \textbf{b}^{(3)}_{6,3,1}, &  \textbf{a}^{(2)}_{6,0,1}}{\tau_A(t)} 
+ 192\,\IEs{\textbf{b}^{(3)}_{6,0,1}, & 1, & \textbf{b}^{(3)}_{6,3,1}, &  \textbf{a}^{(2)}_{6,0,2}}{\tau_A(t)}\nn  \\
& + 192\,\IEs{\textbf{b}^{(3)}_{6,0,1}, & 1, & \textbf{b}^{(3)}_{6,3,2}, &  \textbf{a}^{(2)}_{6,0,1}}{\tau_A(t)} 
- 192\,\IEs{\textbf{b}^{(3)}_{6,0,1}, & 1, & \textbf{b}^{(3)}_{6,3,2}, &  \textbf{a}^{(2)}_{6,0,2}}{\tau_A(t)}\nn  \\
& + 960\,\IEs{\textbf{b}^{(3)}_{6,3,1}, & 1, & \textbf{b}^{(3)}_{6,0,1}, &  \textbf{a}^{(2)}_{6,0,1}}{\tau_A(t)} 
- 960\,\IEs{\textbf{b}^{(3)}_{6,3,1}, & 1, & \textbf{b}^{(3)}_{6,0,1}, &  \textbf{a}^{(2)}_{6,0,2}}{\tau_A(t)}\nn  \\
& + 384\,\IEs{\textbf{b}^{(3)}_{6,3,1}, & 1, & \textbf{b}^{(3)}_{6,3,1}, &  \textbf{a}^{(2)}_{6,0,1}}{\tau_A(t)} 
- 384\,\IEs{\textbf{b}^{(3)}_{6,3,1}, & 1, & \textbf{b}^{(3)}_{6,3,1}, &  \textbf{a}^{(2)}_{6,0,2}}{\tau_A(t)}\nn  \\
& - 384\,\IEs{\textbf{b}^{(3)}_{6,3,1}, & 1, & \textbf{b}^{(3)}_{6,3,2}, &  \textbf{a}^{(2)}_{6,0,1}}{\tau_A(t)} 
+ 384\,\IEs{\textbf{b}^{(3)}_{6,3,1}, & 1, & \textbf{b}^{(3)}_{6,3,2}, &  \textbf{a}^{(2)}_{6,0,2}}{\tau_A(t)}\nn  \\
& - 960\,\IEs{\textbf{b}^{(3)}_{6,3,2}, & 1, & \textbf{b}^{(3)}_{6,0,1}, &  \textbf{a}^{(2)}_{6,0,1}}{\tau_A(t)} 
+ 960\,\IEs{\textbf{b}^{(3)}_{6,3,2}, & 1, & \textbf{b}^{(3)}_{6,0,1}, &  \textbf{a}^{(2)}_{6,0,2}}{\tau_A(t)}\nn  \\
& - 384\,\IEs{\textbf{b}^{(3)}_{6,3,2}, & 1, & \textbf{b}^{(3)}_{6,3,1}, &  \textbf{a}^{(2)}_{6,0,1}}{\tau_A(t)} 
+ 384\,\IEs{\textbf{b}^{(3)}_{6,3,2}, & 1, & \textbf{b}^{(3)}_{6,3,1}, &  \textbf{a}^{(2)}_{6,0,2}}{\tau_A(t)}\nn  \\
& + 384\,\IEs{\textbf{b}^{(3)}_{6,3,2}, & 1, & \textbf{b}^{(3)}_{6,3,2}, &  \textbf{a}^{(2)}_{6,0,1}}{\tau_A(t)} 
- 384\,\IEs{\textbf{b}^{(3)}_{6,3,2}, & 1, & \textbf{b}^{(3)}_{6,3,2}, &  \textbf{a}^{(2)}_{6,0,2}}{\tau_A(t)}\nn  \\
& - \pi^3\left[ 2\,\IEs{\textbf{b}^{(3)}_{6,0,1},&1}{\tau_A(t)} - 4\,\IEs{\textbf{b}^{(3)}_{6,3,1},&1}{\tau_A(t)} 
+ 4\,\IEs{\textbf{b}^{(3)}_{6,3,2},&1}{\tau_A(t)}\right]\nn  \\
& - \frac{12\,\zeta_3}{\pi}\left[ \IEs{\textbf{b}^{(3)}_{6,0,1}}{\tau_A(t)} - 2\,\IEs{\textbf{b}^{(3)}_{6,3,1}}{\tau_A(t)} 
+ 2\,\IEs{\textbf{b}^{(3)}_{6,3,2}}{\tau_A(t)}\right]\nn  \\
& + G\left(0,0,0,0;\frac{1}{t}\right) - G\left(0,0,1,0;\frac{1}{t}\right) - 2\,G\left(0,1,0,0;\frac{1}{t}\right)\nn \\
&+ 2\,G\left(0,1,1,0;\frac{1}{t}\right)
+ G\left(1,0,0,0;\frac{1}{t}\right) - G\left(1,0,1,0;\frac{1}{t}\right)\nn \\
&+ \frac{\pi^2}{6}\left [ G\left(0,0;\frac{1}{t}\right) - 2\,G\left(0,1;\frac{1}{t}\right) 
+ G\left(1,0;\frac{1}{t}\right)\right]\nn \\
& + \frac{16}{3}\zeta_3 G\left(0;\frac{1}{t}\right) - \frac{22}{3}\zeta_3 G\left(1;\frac{1}{t}\right) + \frac{\pi^4}{30},
\end{align}
where the MPLs are written as functions of $1/t$ so that they are real in the region $t\in[t_7,t_8]$.

We finish with a comment on Topology $B$. As already noted, it is obtained from Topology $A$ by  
replacing $t\to1/t$. In eq.~\eqref{eq:allRegA} we defined the vector of functions $f_A(t)$
with fast-converging representations of the master integrals for all real values of $t$. 
We adopt the same conventions for Topology $B$, i.e. all quantities with an index $B$
admit a piecewise definition.
They are simply obtained from those of Topology $A$ using:
\begin{equation}
\vec{f}_B(t)=\vec{f}_A\left(\frac{1}{t}\right)\,.
\end{equation}
Similarly, we have
\begin{equation}
\tau_B(t)=\tau_A\left(\frac{1}{t}\right)\,,\textrm{~~~and~~~}
\mathcal{S}_B(t) =	\mathcal{S}_A\left(\frac{1}{t}\right)\,.
\end{equation}
In particular, with these relations we can obtain Topology $B$ at $t_\text{phys}\in[t_2,t_3]$ from the expressions
given above for Topology $A$ for $1/t_{\textrm{phys}}\in[t_7,t_8]$. As an example, $f_{8,U}^{(2),B}(t_\text{phys})$
is obtained from eq.~\eqref{eq:f8ua}, giving
\begin{align}\begin{split}\label{eq:f8ua}
f_{8,U}^{(2),B}(t_\text{phys}) =\, & 40\, \IEs{1, & \textbf{b}^{(3)}_{6,0,1}, &  \textbf{a}^{(2)}_{6,0,1}}{\tau_B(t_\text{phys})} 
- 40\, \IEs{1, & \textbf{b}^{(3)}_{6,0,1}, &  \textbf{a}^{(2)}_{6,0,2}}{\tau_B(t_\text{phys})} \\
&+ 16 \IEs{1, & \textbf{b}^{(3)}_{6,3,1}, &  \textbf{a}^{(2)}_{6,0,1}}{\tau_B(t_\text{phys})}  
- 16\, \IEs{1, & \textbf{b}^{(3)}_{6,3,1}, &  \textbf{a}^{(2)}_{6,0,2}}{\tau_B(t_\text{phys})} \\
&- 16\, \IEs{1, & \textbf{b}^{(3)}_{6,3,2}, &  \textbf{a}^{(2)}_{6,0,1}}{\tau_B(t_\text{phys})} 
+ 16\, \IEs{1, & \textbf{b}^{(3)}_{6,3,2}, &  \textbf{a}^{(2)}_{6,0,2}}{\tau_B(t_\text{phys})}  \\
& - \frac{i \pi^2}{12}\tau_B(t) + \frac{\zeta_3}{\pi}.
\end{split}\end{align}


\section{Three-loop contributions to the $\rho$ parameter}
\label{Sec8}

We are now ready to give an expression for the three-loop contributions to the $\rho$ parameter
in terms of iterated integrals of modular forms. We start from the expression given in the ancillary
files of ref.~\cite{Grigo:2012ji} renormalised in the $\overline{\textrm{MS}}$ scheme and set the regularisation scale 
$\mu^2 = m_{\textrm{t}}^2$. The result depends on the $SU( N_c)$ colour factors
$C_A= N_c$ and $C_F=( N_c^2-1)/(2 N_c)$, and on the number of
massless quarks $n_l=n_f-2$.\footnote{In the notation
of the ancillary files of ref.~\cite{Grigo:2012ji}, we set
$\texttt{lm}=0$, $\texttt{ca}=\texttt{nc}$ and keep the dependence
on $\texttt{nc}$, $\texttt{cf}$ and $\texttt{nl}$. The SM values correspond to
$\texttt{nc}=3$, $\texttt{cf}=4/3$ and $\texttt{nl}=4$.}

In their expressions, the authors of ref.~\cite{Grigo:2012ji} leave the
elliptic master integrals associated with diagrams $8$, $9$ and $10$ of fig.~\ref{fig:allTopo}
(and their Topology $B$ counterparts) unevaluated, making it particularly convenient to
adapt their expression to our convention. For concreteness, we highlight the two main
changes we make. First, we recall that we write our results in terms of $t = {m_2^2}/{m_1^2}$,
whereas the expression of ref.~\cite{Grigo:2012ji} is written for $x = {m_2}/{m_1}$.
Second, we found it more convenient to compute the masters associated with diagrams $8$, $9$
of fig.~\ref{fig:allTopo} in $d=2$.
Starting from their expression, we thus change variables to $t$ and use the dimension-shifting
relations given in Appendix \ref{app:DRR} to rewrite them in terms of the masters we have computed.
Then, we decompose the three-loop corrections as
\begin{equation}
	\delta^{(2)}(t)=\delta^{(2)}_{\textrm{MPL}}(t)+\delta^{(2)}_{\textrm{ell}}(t)
\end{equation}
where $\delta^{(2)}_{\textrm{ell}}(t)$ contains all contributions related to the elliptic master integrals
$f_8^{(2),A}$, $f_8^{(2),B}$, $f_9^{(2),A}$, $f_9^{(2),B}$, $f_{10}^{A}$ and $f_{10}^{B}$.
The remaining contribution $\delta^{(2)}_{\textrm{MPL}}(t)$ only depends on MPLs and we do not
discuss it further in this paper.

The elliptic component $\delta^{(2)}_{\textrm{ell}}(t)$ can be written in a very compact form. For that, we define
the vector of coefficients
\begin{align}
\mathcal{C}^A(t)=C_F  N_c\begin{pmatrix}
 \frac{(2C_F- N_c)(t-11) (t-9) t}{144 (t-1)}+\frac{(t-9) t \left(5 t^2-28 t-9\right)}{108 (t-1)^2}\\
 -\frac{(2C_F- N_c)(t-13) (t-9) (t-3) t}{144 (t-1)}-\frac{(t-9) t \left(7 t^2-36 t-27\right)}{108 (t-1)}\\
 \frac{2C_F- N_c}{12}
\end{pmatrix}\,,
\end{align}
and its counterpart
\begin{equation}
	\mathcal{C}^B(t)=\mathcal{C}^A\left(\frac{1}{t}\right)\,.
\end{equation}
We then find
\begin{equation}
	\delta^{(2)}_{\textrm{ell}}(t)=\left(\mathcal{C}^A(t )\right)^T\mathcal{S}^A(t ) 
	\left(\begin{array}{c} f_{8,U}^{(2),A}(t ) \\ f_{9,U}^{(2),A}(t )   \\ f^{A}_{10,U}(t ) \end{array}\right)
	+t\left(\mathcal{C}^B(t )\right)^T\mathcal{S}^B(t ) 
	\left(\begin{array}{c} f_{8,U}^{(2),B}(t ) \\ f_{9,U}^{(2),B}(t )   \\ f^{B}_{10,U}(t ) \end{array}\right)\,.
\end{equation}
More explicitly,
\begin{align}
	&\frac{\delta^{(2)}_{\textrm{ell}}(t)}{C_FN_c}=
	\frac{1}{12} (2 C_f-N_c)\left(f^{A}_{10,U}(t)
	+t\,f^{B}_{10,U}(t)\right)\nn\\
	&- \left(\frac{(2C_F- N_c)(t-13)(t-3)}{6 (t-1)^2}+
	\frac{ 2\left(7 t^2-36 t-27\right)}{9 (t-1)^2}\right)
	\frac{f_{9,U}^{(2),A}(t )}{\Psi^A_1(t)}\nn\\
	&+\left(\frac{(2C_F- N_c)(t-11) (t-9) t}{144 (t-1)}+\frac{(t-9) t \left(5 t^2-28 t-9\right)}{108 (t-1)^2}\right)\Psi^A_1(t)f_{8,U}^{(2),A}(t)\nn\\
	&+\left(\frac{(2C_F- N_c)(t-13) (t-9) (t-3) t}{144 (t-1)}
	+\frac{(t-9) t \left(7 t^2-36 t-27\right)}{108 (t-1)}\right)
	\Phi^A_1(t)f_{8,U}^{(2),A}(t)\nn\\
	&+\left(
	\frac{(2 C_F-N_c)(3 t-1)  (13 t-1)t}{6 (t-1)^2 }-
	\frac{ 2\left(27 t^2+36 t-7\right)t}{9 (t-1)^2}
	\right)\frac{f_{9,U}^{(2),B}(t )}{\Psi^B_1(t)}\\
	&-\left(
	\frac{(2 C_F-N_C)(9 t-1) (11 t-1) }{144 (t-1)t}
	-\frac{(9 t-1) \left(9 t^2+28 t-5\right)}{108 (t-1)^2 t}
	\right)\Psi^B_1(t)f_{8,U}^{(2),B}(t)\nn\\
	&+\left(
	\frac{(2 C_F-N_c)(3 t-1) (9 t-1) (13 t-1) }{144 (t-1) t^2}
	-\frac{(9 t-1) \left(27 t^2+36 t-7\right)}{108 (t-1) t^2}
	\right)\Phi^B_1(t)f_{8,U}^{(2),B}(t)\,.\nn
\end{align}

\begin{figure}
\centering
\includegraphics[width =9cm]{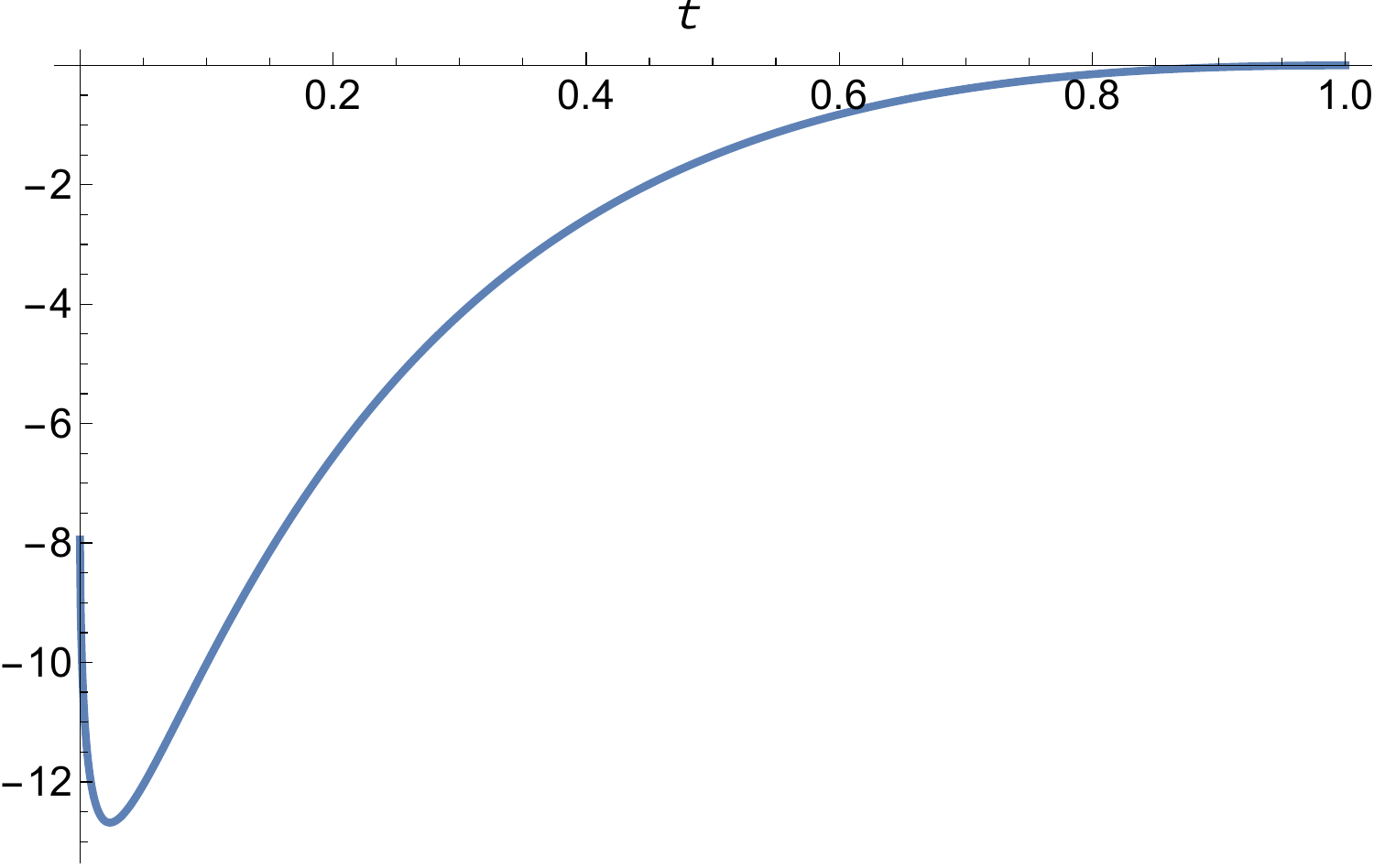}
\caption{Three-loop corrections to the $\rho$ parameter, for $N_c=3$, 
$C_F=4/3$ and $n_l=4$ massless quarks, as a function of $t=m_2^2/m_1^2$ 
which we vary between 0 and 1.}
\label{fig:plot} 
\end{figure}

We provide a set of ancillary files that allows to evaluate $\delta^{(2)}(t)$ from our 
expression in terms of iterated integrals of modular forms. 
Using the \texttt{MATHEMATICA} script we provide, we find 
\begin{equation}
	\delta^{(2)}(t_\text{phys})=-9.03594\ldots
\end{equation}
for $t_\text{phys}=5\cdot10^{-4}$, $N_c=3$ and $n_l=4$. Using the same scripts,
we plotted $\delta^{(2)}(t)$ for $t\in[0,1]$ in fig.~\ref{fig:plot}.
We find complete agreement with the values
results of \cite{Chetyrkin:1995ix,Grigo:2012ji,Ablinger:2017bjx,Blumlein:2018aeq}.

\section{Conclusion}
\label{sec:conclusion}

In this paper we have presented for the first time fully analytic results in terms of eMPLs and iterated
Eisenstein integrals for the
three-loop corrections to the $\rho$ parameter in the SM with two massive quark flavours.
This computation was originally considered as an expansion in the ratio of the quark masses in ref.~\cite{Grigo:2012ji}.
An important ingredient in our calculation is the realisation that the homogeneous
second-order differential operator appearing in this computation is identical
to the differential operator that appears in the computation of the well-known sunrise graph.
As a consequence, all integrals can be expressed
in terms of the exact same class of functions as the sunrise graph, which are also well-studied
functions in pure mathematics. We can draw upon the
knowledge of these functions to analytically continue them using tools and algorithms
developed for the sunrise graph.
This distinguishes our computation from the results of refs.~\cite{Ablinger:2017bjx,Blumlein:2018aeq}, where
a novel class of special functions was introduced for the same class of integrals.

Besides ref.~\cite{Honemann:2018mrb}, our computation is only the second time that iterated integrals of modular forms
have been used to obtain fully analytic results for a complete physical observable.
We believe that the techniques that we have used in our computation can have an impact also on the computation
of other physical observables, and that they pave the way for obtaining more
results involving this class of special functions.

\section*{Acknowledgments}
We are grateful to Lorenzo Tancredi for discussion. This work was supported in part by the ERC starting grants  637019 ``MathAm'' (CD), and the grants ``2LoopAmps4LHC'' (SA) and ``ElliptHiggs'' of the Fonds National de le Recherche Scientifique (FNRS), Belgium.

\appendix


\section{Differential equations for the non-elliptic master integrals}
\label{app:DEQ}

As written in eq.~\eqref{f1f7DE}, the differential equations 
for the master integrals which are expressible in terms of MPLs, i.e.~$f_1$ 
through~$f_7$ of eq.~\eqref{eq:masters}, is given by
 \begin{equation}
 \partial_t f_a = \epsilon \,\frac{(A_0)_{ak}}{t} f_k + \epsilon \frac{(A_1)_{ak} }{t - 1}f_k\,, \qquad 1\le a,k\le 7\,.
 \end{equation}
The matrices $A_0$ and $A_1$ are given by
\be
A_0 = \left(\begin{array}{ccccccc}
-1 & 0 & 0 & 0 & 0 & 0 & 0 \\
 0 & 0 & 0 & 0 & 0 & 0 & 0 \\
 1 & 0 & 0 & 0 & 0 & 0 & 0 \\
 0 & 0 & 0 & 1 & 2 & 0 & 0 \\
 0 & 0 & 0 & -1 & -2 & 0 & 0 \\
 0 & 0 & 0 & 0 & 0 & 0 & 0 \\
 0 & 0 & 0 & 0 & 0 & 0 & 1 \\
\end{array}
\right),
\textrm{~~~
and 
~~~}
A_1 = \left(\begin{array}{ccccccc}
 0 & 0 & 0 & 0 & 0 & 0 & 0 \\
 0 & 0 & 0 & 0 & 0 & 0 & 0 \\
 -1 & 0 & -2 & 0 & 0 & 0 & 0 \\
 0 & 0 & 0 & -4 & -2 & 0 & 0 \\
 0 & 0 & 0 & 0 & 0 & 0 & 0 \\
 0 & 0 & 0 & 0 & 0 & 0 & 0 \\
 2 & 0 & 2 & 0 & 0 & -6 & -4 \\
\end{array}
\right).
\ee


\section{Dimension-shift identities for elliptic master integrals}
\label{app:DRR}

In this appendix we give the dimensional recurrence relations for the master integrals $f_8^{(2)}$ and $f_9^{(2)}$. We denote $f_8^{(d)}$ and $f_9^{(d)}$ the master integrals in $d$ dimensions. Then, the full dimensional recurrence relations are given by:

\begin{align}\label{eq:f8d4e}
f_8^{(4 - 2\epsilon)} = \,\,\, & \frac{1}{6 \, \epsilon^3 (1 - 3\,\epsilon + 2\,\epsilon^2)(2 - 9\,\eps + 9\,\epsilon^2)}\left[ 3\left(6\,t -21 \,\epsilon\,t - \epsilon\,t^2\right)f_1^{(4-2\epsilon)} \right. \nn \\
& \left. + \left(12 - 6\,t - 45\,\epsilon + 21\,\epsilon\,t + 2\,\epsilon\,t^2\right)f_2^{(4-2\epsilon)} + \left(27\,t - 21\,t^2 - 7\,t^3 + t^4\right)f_9^{(2 - 2\epsilon)}  \right. \nn \\
& \left. + \left(10\,t^2 - 9\,t - t^3 - 27\,\epsilon + 108\,\epsilon\,t - 15\,\epsilon\,t^2 - 2\,\epsilon\,t^3\right)f_8^{(2 - 2\epsilon)}\right],
\end{align}

\begin{align}\label{eq:f9d4e}
f_9^{(4 - 2\epsilon)} = \,\,\, & \frac{1}{6\, \epsilon^3\,(-1 + 3\,\epsilon)(1 - 3\,\epsilon + 2\,\epsilon^2)}\left[3\left(3 - 9\,\epsilon - \epsilon\,t\right)f_1^{(4-2\epsilon)} + \left(6\,\epsilon + 2\,\epsilon\,t - 3\right)f_2^{(4-2\epsilon)} \right. \nn \\
& \left. + \left(10\,t - 9 - t^2 + 8\,\epsilon - 2\,\epsilon\,t^2\right)f_8^{(2 - 2\epsilon)} + \left(9\,t - 10\,t^2 + t^3\right)f_9^{(2 - 2\epsilon)}\right].
\end{align}

By expanding in $\epsilon$ the right-hand side of eqs.~\eqref{eq:f8d4e} and \eqref{eq:f9d4e}, 
it is possible to show that the poles of $f_8^{(4 - 2\epsilon)}$ and $f_9^{(4 - 2\epsilon)}$ do 
not depend on $f_8^{(2)}$ and $f_9^{(2)}$. The elliptic contributions appear starting from order $\epsilon^0$:

\begin{align}\label{eq:f8d4}
f_8^{(4 - 2\epsilon)} = \,\,\, & \frac{1 + t}{\epsilon^3} + \frac{45 + 48\,t - t^2 - 18\, \log t}{12\epsilon^2} \nn \\
& + \frac{65 + 2\,\pi^2 + 80\,t + 2\,\pi^2\,t - 5\,t^2 - 48\,t\,\log t + 2\,t^2\,\log t + 6\,t\,\log^2 t}{8\,\epsilon} \nn \\
& + \frac{1}{48}\left[405 + 45\,\pi^2 + 840\,t + 48\,\pi^2\,t - 145\,t^2 - \pi^2\,t^2 - 720\,t\,\log t - 18\,\pi^2\,t\log t \right. \nn \\
& \left. + 90\,t^2\,\log t + 144\,t\,\log^2 t - 6\,t^2\,\log^2 t - 6\,t\,\log^3 t - 48\zeta_3 - 48\,t\,\zeta_3 \right. \nn \\
& \left. + \left(40\,t^2 - 36\,t - 4\,t^3\right)f_8^{(2)} + \left(108\,t - 84\,t^2 - 28\,t^3 + 4\,t^4\right)f_9^{(2)}\right] + O(\epsilon),
\end{align}

\begin{align}\label{eq:f9d4}
f_9^{(4 - 2\epsilon)} = \,\,\, & - \frac{1}{\epsilon^3} + \frac{t - 15 + 9\,\log t}{6\,\epsilon^2} + \frac{4\,t - 16 - \pi^2 + 18\,\log t - 2\,t\,\log t - 3\,\log^2 t}{4\,\epsilon} \nn \\
& + \frac{1}{24}\left[100\,t - 60 - 15\,\pi^2 + \pi^2\,t + 252\,\log t + 9\,\pi^2\,\log t - 72\,t\,\log t - 54\,\log^2 t \right. \nn \\
& \left. + 6\,t\,\log^2 t + 6\,\log^3 t + 24\,\zeta_3 + \left(36 - 40\,t + 4\,t^2\right)f_8^{(2)} + \left(36\,t + 40\,t^2 - 4\,t^3\right)f_9^{(2)}\right] \nn \\
& + O(\epsilon).
\end{align}



\bibliographystyle{JHEP}
\bibliography{biblio}

\end{document}